\documentclass[aps,pra,twocolumn,floatfix,showpacs,showkeys,amsmath,amssymb]{revtex4}
\usepackage{graphicx}
\usepackage[hypertex]{hyperref}

\begin{document}

\newlength\smallfigwidth
\smallfigwidth=3.2 in

\preprint{Submitted to PRA}

\title{Electromagnetic modes in dielectric equilateral triangle resonators}
\author{G.\ M.\  Wysin}
\email{wysin@phys.ksu.edu}
\homepage{http://www.phys.ksu.edu/~wysin}
\affiliation{
Department of Physics \\
Kansas State University \\
Manhattan, KS 66506-2601
}
\date{July 1, 2005}

\begin{abstract} 
Resonant electromagnetic modes are analyzed inside a dielectric cavity 
of equilateral triangular cross section and refractive index $\mathsf{n}$,
surrounded by a uniform medium of refractive index $\mathsf{n'}$.
The field confinement is determined only under the requirements needed
to maintain total internal reflection of the internal electromagnetic fields, 
matched to exponentially decaying evanescent waves outside the cavity.
Two-dimensional electromagnetics is considered,  with no dependence on the 
coordinate perpendicular to the cross section;  hence, independent
transverse electric (TE) and transverse magnetic (TM) polarizations 
are described separately.
A linear combination of six plane waves is sufficient within the cavity,
whose wavevectors are related by $120^{\circ}$ rotations and whose phases
are related by Fresnel reflection coefficients.
Generally, the mode spectrum becomes sparse and the minimum mode frequency 
increases rapidly as the index ratio $\mathsf{N=n/n'}$ approaches 2.
For specified quantum numbers and $\mathsf{N}$, the TM modes are lower 
in frequency than the TE modes.
Assuming the evanescent boundary waves escape at the triangle vertices,
TE modes generally are found to have greater confinement of the fields
inside the cavity and much higher quality factors than TM modes.
\end{abstract} 

\maketitle

\section{Introduction}
Confinement of electromagnetic fields only by total internal reflection (TIR)
offers a path for design of simple optical resonators or lasers with low volume 
and low threshold power.
To this end, various two-dimensional (2D) geometries have been considered 
for their different influence on the mode properties, including 
disks,\cite{McCall92} cleaved\cite{Chang00} and etched\cite{Lu04} 
semiconductor triangles, squares,\cite{Poon01,Fong03,Moon03} and zeolite 
ALPO$_4$-5 hexagonal crystals.\cite{Vietze98,Braun00}
The problem of equilateral triangle resonators (ETR) considered here
is particularly interesting, because the symmetry present allows for a 
considerable simplification of the analysis.
Good estimates for the ETR mode wavefunctions, frequencies, and lifetimes
(or quality factors) are developed here, based on a set of six plane waves
within the resonator, matched to each other by Fresnel factors,
producing exterior evanescent waves.

Chang \textit{et al.}\cite{Chang00} analyzed the ETR wavefunctions under the
assumption of Dirichlet boundary conditions (DBC), using an exactly known
solution\cite{Lame52}, for transverse magnetic (TM) polarization only.
In the presence of a dielectric--dielectric boundary, however, the
fields do not go to zero on the boundary, instead, the field confinement can 
be provided by TIR, assuming a refractive index in the cavity, $\mathsf{n}$,
greater than that of the surroundings, $\mathsf{n'}$.
Huang \textit{et al.}\cite{Huang99,Guo00,Huang01} made a more complete analysis
for an ETR with dielectric boundary conditions, approximately matching 
interior standing wave fields to exterior evanescent waves.
Mode frequencies and quality factors $Q$ were estimated\cite{Guo00}
from 1000 to as high as 20000, using the finite-difference 
time-domain technique (FDTD) and Pad\'e approximation\cite{Dey98},
with the highest $Q$ values associated with TM polarization.
On the other hand, typical $Q$ values from 20 -- 150 were measured in 
photoluminescence (PL) experiments\cite{Lu04} on GaInAs-InP ETRs
with edges from 5 -- 20 $\mu$m. 
Recently I analyzed the ETR modes using the DBC 
approximation\cite{Wysin05a}, and assuming the escape of evanescent
boundary waves at the triangle vertices as the primary decay mechanism, 
estimated typical $Q$'s from 10 -- 500, with the highest values for TE 
polarization.
Both of these theoretical approaches involve approximations, thus, it is 
important to consider an alternative description of the modes and clarify
how the polarization influences the mode lifetimes.

The DBC approximation applied previously\cite{Chang00,Wysin05a} clearly does 
not describe the fields correctly at a dielectric--dielectric boundary.
The goal here is to use the general knowledge of the modes from the
DBC analysis, but do the correct matching of plane wave fields inside
the cavity with evanescent fields on the outside.
This matching is accomplished by employing the Fresnel reflection
coefficients correctly for all the plane wave components present in
the cavity.
It is assumed that all the electromagnetic field components do not
depend on a $z$-coordinate perpendicular to the triangular cross section
(i.e., longitudinal wavevector $k_z=0$).
Then Maxwell's equations and associated dielectric boundary 
conditions lead to separated problems for TM and TE modes.
Each polarization is controlled by one component of the electromagnetic
field that must be continuous across the dielectric-dielectric boundary.
For the TM modes, the controlling wavefunction is the longitudinal electric 
field, $\psi=E_z$; for TE modes, it is the longitudinal magnetic field, 
$\psi=H_z$.

The exact solution for an ETR with Dirichlet boundary conditions is
a superposition of six plane waves of equal strengths but different 
phases and wavevectors.
The waves undergo a $-\pi$ phase shift when reflecting from the faces,
as required such that the incident and reflected parts cancel exactly
at the boundary.
Any one of these waves, when followed through a sequence of reflections 
due to TIR, returns to its original direction after six reflections
(see Fig.\ 1 of Ref.\ \onlinecite{Wysin05a}).

The situation is similar for the ETR with \textit{dielectric boundary
conditions} or \textit{Maxwell boundary conditions} (MBC), where
the correct field matching for Maxwell's equations must be applied.
For either TE or TM polarization, I show that a combination of six plane 
waves is still needed within the cavity, related to each other by the 
reflections from the three cavity faces.
For a resonant mode whose fields are TIR-confined within the cavity,
all the waves must impinge on the faces at incident angles greater than
the critical angle $\theta_c$, where 
\begin{equation}
\label{crit}
\sin\theta_c = \frac{1}{\mathsf{N}}, \qquad  \mathsf{N=\frac{n}{n'}}.
\end{equation}
The generalization here compared to the DBC problem, is that when reflected 
from the faces, the waves undergo phase shifts determined by unit-modulus
Fresnel coefficients, as well as being rotated in propagation direction.
The goal here is to determine the correct wavevectors and complex amplitudes 
of these six plane waves such that a fully self-consistent wavefunction is 
determined.

When incident on the faces from inside the cavity, the waves
produce evanescent waves in the exterior region just outside the cavity.
It is difficult to give an exact description of the exterior fields, however,
an approximate description is possible provided the six interior plane
waves are incident on the cavity faces at angles sufficiently greater
than $\theta_c$.
In that case, the penetration depth into the exterior region 
$d=\vert k'_{\perp}\vert^{-1}$, is much less than the cavity size or edge $a$.
Then, there are only strong evanescent fields very close to the cavity,
and a description of these based on the transmission amplitude via Fresnel's 
equations is appropriate.
The mode lifetime and $Q$ are estimated based on the assumption that 
the evanescent boundary waves radiate to the cavity surroundings when they 
reach the triangle vertices.
This is a ``strong damping'' approximation, in the sense that the reflection
or radiation of boundary wave energy back into the cavity is assumed to be 
insignificant.

The presentation proceeds as follows.
First, the MBC field matching of the six plane waves is described.
Equations for allowed quantum indexes are found, giving
solutions for the mode wavevectors and wavefunctions.
The wavefunction description is general enough that it describes equally
well the DBC, TM and TE problems, or any other boundary condition
whose reflection phase shift is determined by incident angle.
Lifetimes will be estimated using the boundary wave radiation, and
compared to the simpler DBC theory and other analysis.

\section{EM fields description}
\subsection{Six interior plane waves}
\begin{figure}
\includegraphics[angle=0.0,width=\columnwidth]{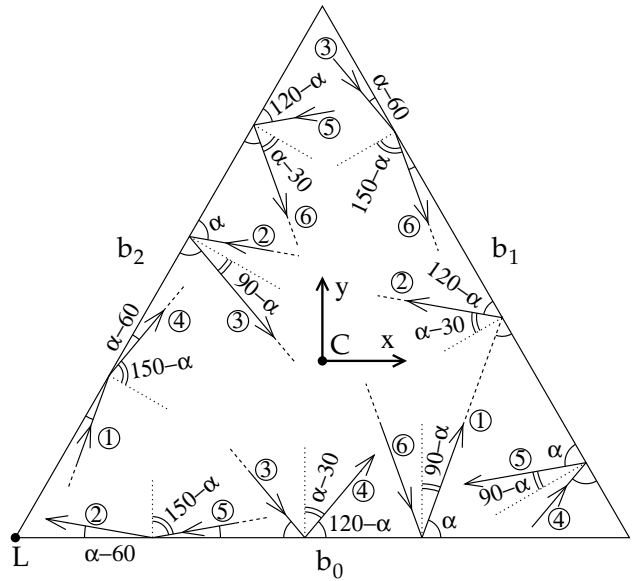}
\caption{ \label{reflections} $xy$ coordinates, with origin
at triangle center, and boundaries $b_0$, $b_1$ and $b_2$.  Rays 
represent the six planes waves propagating within the cavity,   
sketched for the value, $\alpha=70^{\circ}$.  The plane wave incident 
angles here determine the reflection amplitudes and phases used in 
Table \ref{matching}.
}
\end{figure}

Here I describe the notation for the interior waves ($\psi$), with assumed 
frequency $\omega=c^{*}k$, where $c^{*}=c/\sqrt{\epsilon\mu}$ is the light 
speed in the cavity, and the wavevector modulus $k$ is to be determined.
The ETR has edge length $a$, and $xy$ coordinates are used where the
origin is set at its geometrical center, Fig.\ \ref{reflections}.
An initial wave \textcircled{$\scriptstyle 1$} is supposed to emerge 
from lower edge $b_0$ at an angle $\alpha_1=\alpha$ relative to the $x$-axis, 
and hence has wavevector components
$\vec{k}_1 \equiv (k_x,k_y)=(k\cos\alpha, k\sin\alpha)$.
Only certain values of $\alpha$ will lead to a solution for an eigenmode
of the cavity; determination of the possible values of $\alpha$ is an
essential part of the solution presented here.

This initial wave \textcircled{$\scriptstyle 1$} was formed by reflection of
a wave incident on boundary $b_0$ at incident angle 
$\theta_{i,1}=90^{\circ}-\alpha$.
By consideration of the triangular geometry and using the law of reflection,
it is seen that the incident angle when impinging on boundary $b_1$ will be 
$\theta_{i,2}=60^{\circ}-\theta_{i,1}$, see Fig.\ \ref{reflections}.
As the wave subsequently reflects from the boundaries in ordered sequence
$b_1$, $b_2$, $b_0$, etc., the incident angles on each succeeding boundary
simply oscillate between only two values symmetrically above and below 
$60^{\circ}$.
Thus, without loss of generality, one can assume $\alpha\ge 60^{\circ}$
for the remaining analysis, and take $k_x$ and $k_y$ of wave 
\textcircled{$\scriptstyle 1$} as both positive.
The sequence of waves generated from wave \textcircled{$\scriptstyle 1$}, 
by sequential reflections from $b_1$, $b_2$, $b_0$, $b_1$ and so on, are 
labeled as waves \textcircled{$\scriptstyle 2$}, \textcircled{$\scriptstyle 3$},
\textcircled{$\scriptstyle 4$}, \textcircled{$\scriptstyle 5$} and 
\textcircled{$\scriptstyle 6$}.
Finally,  wave \textcircled{$\scriptstyle 6$} emerges from $b_2$ and 
impinges on $b_0$ to regenerate an amplitude of wave 
\textcircled{$\scriptstyle 1$}.
To produce a consistent solution, this regeneration of wave 
\textcircled{$\scriptstyle 1$} must be in phase with the original wave 
\textcircled{$\scriptstyle 1$}.

Considerations of the law of reflection together with the equilateral
geometry leads to the basic properties of the six waves, as summarized in 
Table \ref{6waves}.
The $xy$ components of each wavevector are defined from the $\alpha_l$ in
the usual way,
\begin{equation}
\label{kcomps}
\vec{k}_l = k(\cos\alpha_l, \sin\alpha_l).
\end{equation}
It is seen that pairs of wavevectors, $[\vec{k}_1, \vec{k}_6]$,
and $[\vec{k}_2, \vec{k}_3]$ and $[\vec{k}_4, \vec{k}_5]$, are related
to each other by changing $\alpha\to -\alpha$.
Each pair is related to the others by $120^{\circ}$ counterclockwise rotations 
around the $\hat{z}$-axis, denoted by operator $R$, which can be represented 
by the square matrix,
\begin{equation}
\label{Rdef}
R = \begin{pmatrix} -1/2 & -\sqrt{3}/2 \\
                    \sqrt{3}/2 & -1/2 
\end{pmatrix} .
\end{equation}
The pair $[\vec{k}_1, \vec{k}_6]$ can be considered the original waves,
from which the others are obtained.
Wavevector $\vec{k}_6$ is simply obtained from $\vec{k}_1$ by reflection
across the $\hat{x}$-axis.
%

Due to the triangular symmetry, and the unit-modulus reflection coefficient
under TIR, all six waves must have equal magnitudes, but different phases.
Then the net wavefunction within the cavity is written as a sum over
the waves
\begin{equation}
\label{psi-six}
\psi= \sum_{l=1}^{6} A_l e^{i\vec{k}_l\cdot\vec{r}}
\end{equation}
where the $A_l$ are all of unit modulus, and $\psi=H_z$ for TE modes
or $\psi=E_z$ for TM modes.
In both polarizations, $\psi$ must be continuous across the
cavity boundary.
The TE and TM wavefunctions will not be the same, however, due to 
the different phase shifts implied by the Fresnel factors.

\begin{table}
\caption{\label{6waves} Definitions of the parameters of the plane waves
labeled by wavevectors $\vec{k}_l$, within the triangular cavity.  
$\alpha_l$ is the angle that each $\vec{k}_l$ makes to the $x$-axis.
$R$ is the operator for rotation through $+120^{\circ}$ around the 
$\hat{z}$-axis. 
}
\begin{ruledtabular}
\begin{tabular}{lcccccc}
   & \textcircled{$\scriptstyle 1$} & \textcircled{$\scriptstyle 2$} 
   & \textcircled{$\scriptstyle 3$} & \textcircled{$\scriptstyle 4$} 
   & \textcircled{$\scriptstyle 5$} & \textcircled{$\scriptstyle 6$} \\
\hline
$\alpha_l=$ & $\alpha$ & $-\alpha+240^{\circ}$ & $\alpha+240^{\circ}$ &
             $-\alpha+120^{\circ}$  & $\alpha+120^{\circ}$ & $-\alpha$ \\
$\vec{k}_l=$ & $\vec{k}_1$ & $R^2\cdot\vec{k}_6$ & $R^2\cdot\vec{k}_1$ &
               $R\cdot\vec{k}_6$ & $R\cdot\vec{k}_1$ & $\vec{k}_1(-\alpha)$ \\
\end{tabular}
\end{ruledtabular}
\end{table}

Now consider what happens to these waves in terms of their interaction 
with the lower boundary, $b_0$, which lies parallel to $\hat{x}$ at 
$y=-a/(2\sqrt{3})$. 
By appropriate symmetry transformations, the effects on the other two 
boundaries can be inferred.
Three of the waves must be receding from $b_0$ and three must be approaching
$b_0$.
Using the fact that $60^{\circ} \le \alpha < 90^{\circ}$, for example, writing
$\alpha=60^{\circ}+\vartheta$, where $\vartheta<30^{\circ}$, the respective angles
of the $\vec{k}_l$ to the $\hat{x}$-axis are 
$60^{\circ}-\vartheta$, 
$180^{\circ}-\vartheta$, 
$-60^{\circ}+\vartheta$,
$60^{\circ}-\vartheta$,
$180^{\circ}+\vartheta$,
and $-60^{\circ}-\vartheta$.
Those that fall in the range from $0^{\circ}$ to $-180^{\circ}$ (including
shifts by multiples of $360^{\circ}$) are incident on $b_0$.
Thus, the waves \textcircled{$\scriptstyle 3$}, \textcircled{$\scriptstyle 5$}
and \textcircled{$\scriptstyle 6$} are incident on $b_0$.
Their incident angles are obtained from the formula, 
$\theta_i=90^{\circ}+\alpha_l$, as $-90^{\circ}$ is the outward normal 
direction to $b_0$.
Then waves \textcircled{$\scriptstyle 3$}, \textcircled{$\scriptstyle 5$} 
and \textcircled{$\scriptstyle 6$} have incident angles on $b_0$,
\begin{equation}
\label{inc356}
\theta_{i,3}=\alpha-30^{\circ}, \quad
\theta_{i,5}=\alpha-150^{\circ}, \quad
\theta_{i,6}=90^{\circ}-\alpha.
\end{equation}
One sees that $\theta_{i,6}$ always has the smallest magnitude.
$\theta_{i,5}$ is less than zero because wave \textcircled{$\scriptstyle 5$} 
has a negative wavevector component along $\hat{x}$.

Extending such arguments to the other faces, the ray diagram of Fig.\ 
\ref{reflections} summarizes the results, showing the incident angles of 
all the waves on each of the faces.
It is important to notice that the only incident angles in the problem
are the three of Eq.\ \ref{inc356}.
For example, wave \textcircled{$\scriptstyle 1$} is incident on 
$b_1$, but at incident angle $\theta_{i,3}$.
The symmetries represented in Fig.\ \ref{reflections} will be exploited
to determine the correct wave amplitudes.

\begin{table}
\caption{\label{matching} Relations between the incident and reflected
wave amplitudes on the lower boundary ($b_0$), the upper right boundary ($b_1$)
and the upper left boundary ($b_2$).  The net reflection phase shifts
are $\Delta_l = \delta_l-k_{ly}\frac{a}{\sqrt{3}}$, where 
$\delta_l=\delta(\theta_{i,l})$ is the Fresnel reflection phase shift.
See Fig.\ \ref{reflections} for the geometrical reasoning behind this table.
}
\begin{ruledtabular}
\begin{tabular}{cccc}
 boundary  & incident &  $\theta_i$  & reflected  \\
\hline
$b_0$  & \textcircled{$\scriptstyle 3$}, $A_3$
       & $\theta_{i,3}=\alpha-30^{\circ}$
       & \textcircled{$\scriptstyle 4$}, $A_4 = A_3 e^{i\Delta_3}$ \\
$b_0$  & \textcircled{$\scriptstyle 5$}, $A_5$
       & $\theta_{i,5}=\alpha-150^{\circ}$
       & \textcircled{$\scriptstyle 2$}, $A_2 = A_5 e^{i\Delta_5}$ \\
$b_0$  & \textcircled{$\scriptstyle 6$}, $A_6$
       & $\theta_{i,6}=90^{\circ}-\alpha$
       & \textcircled{$\scriptstyle 1$}, $A_1 = A_6 e^{i\Delta_6}$ \\
\hline
$b_1$  & \textcircled{$\scriptstyle 1$}, $A_1$
       & $\theta_{i,3}$
       & \textcircled{$\scriptstyle 2$}, $A_2 = A_1 e^{i\Delta_3}$ \\
$b_1$  & \textcircled{$\scriptstyle 3$}, $A_3$
       & $\theta_{i,5}$
       & \textcircled{$\scriptstyle 6$}, $A_6 = A_3 e^{i\Delta_5}$ \\
$b_1$  & \textcircled{$\scriptstyle 4$}, $A_4$
       & $\theta_{i,6}$
       & \textcircled{$\scriptstyle 5$}, $A_5 = A_4 e^{i\Delta_6}$ \\
\hline
$b_2$  & \textcircled{$\scriptstyle 5$}, $A_5$
       & $\theta_{i,3}$
       & \textcircled{$\scriptstyle 6$}, $A_6 = A_5 e^{i\Delta_3}$ \\
$b_2$  & \textcircled{$\scriptstyle 1$}, $A_1$
       & $\theta_{i,5}$
       & \textcircled{$\scriptstyle 4$}, $A_4 = A_1 e^{i\Delta_5}$ \\
$b_2$  & \textcircled{$\scriptstyle 2$}, $A_2$
       & $\theta_{i,6}$
       & \textcircled{$\scriptstyle 3$}, $A_3 = A_2 e^{i\Delta_6}$ \\
\end{tabular}
\end{ruledtabular}
\end{table}

\subsection{Field matching at the dielectric interface -- 
Maxwell boundary conditions}
\begin{figure}
\includegraphics[angle=-90.0,width=\columnwidth]{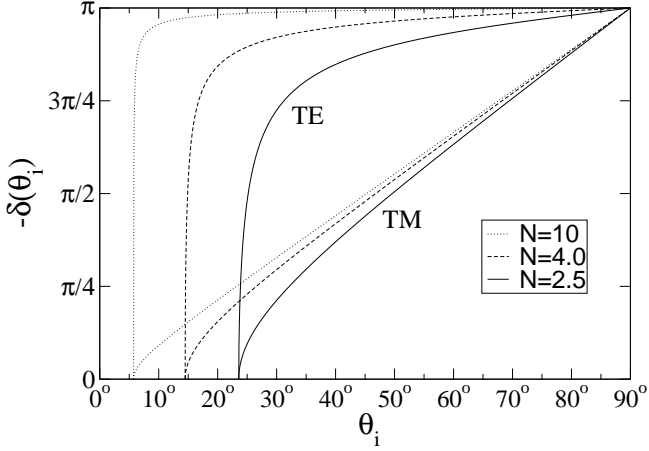}
\caption{ \label{delta} Dependence of the TIR Fresnel reflection phase shifts
on incident angle, for TM [Eq.\ (\ref{deltaTM})] and TE [Eq.\ (\ref{deltaTE})]
polarization at the indicated refractive index ratios $\mathsf{N}$, assuming 
unit magnetic permeabilities.
}
\end{figure}

Now consider the reflections of the waves \textcircled{$\scriptstyle 3$}, 
\textcircled{$\scriptstyle 5$}, and \textcircled{$\scriptstyle 6$} from 
$b_0$, starting with wave \textcircled{$\scriptstyle 6$}.
When wave \textcircled{$\scriptstyle 6$} impinges on $b_0$, its reflection 
regenerates wave \textcircled{$\scriptstyle 1$} and it also produces an 
exterior evanescent wave, denoted \textcircled{$\scriptstyle 6'$}, of amplitude 
$F_6$ (measured on the boundary) and wavevector $\vec{k}_6'$, with 
$k_{6',x}=k_{6,x}$ by Snell's Law.
Ignoring the other interior waves, we match incident wave 
\textcircled{$\scriptstyle 6$} and reflected wave \textcircled{$\scriptstyle 1$} 
to the exterior evanescent wave, \textcircled{$\scriptstyle 6'$}.
Matching the net $\psi$ at $y=-a/(2\sqrt{3})$, we have terms all 
proportional to the common phase factor, $e^{i(k_x x-\omega t)}$, where 
$k_x=k\cos\alpha$,
\begin{equation}
\label{match61}
A_6 e^{i k_{6y}\cdot(\frac{-a}{2\sqrt{3}})} +
A_1 e^{i k_{1y}\cdot(\frac{-a}{2\sqrt{3}})} = F_6 .
\end{equation} 
On the other hand, in the usual analysis of the reflection, the 
correct boundary conditions for Maxwell's equations will relate the 
incident ($\psi_{\mathrm{inc.}}$) and reflected ($\psi_{\mathrm{refl.}}$)
amplitudes, \textit{measured at the boundary}, by a Fresnel reflection
 factor $r(\theta_i)$,
\begin{equation}
\label{Fres-def}
\frac{\psi_{\mathrm{refl.}}}{\psi_{\mathrm{inc.}}} 
   = r(\theta_i)=e^{i\delta(\theta_i)}. 
\end{equation}
Here $\delta(\theta_i)$ is the phase shift experienced by the wave 
upon reflection, for incident angle $\theta_i$, in conditions of TIR.
The $\theta_i$-dependence of the different Fresnel factors appearing for 
TE and TM polarization is discussed subsequently in Sec.\ \ref{solve-omega},
see Fig.\ \ref{delta} for typical dependence with dielectric boundary 
conditions.

Expressions (\ref{match61}) and (\ref{Fres-def}) imply that the wave 
amplitudes $A_6$ and $A_1$ have a relation,
\begin{equation}
r_6 = e^{i\delta_6} = \frac{A_1 e^{-i k_{1y} \frac{a}{2\sqrt{3}} } }
                           {A_6 e^{-i k_{6y} \frac{a}{2\sqrt{3}} } }
= \frac{A_1}{A_6} e^{i k_{6y} \frac{a}{\sqrt{3}} }.
\end{equation}
The phase factor involves the y-component of $\vec{k}_6$, which is
the negative of the y-component of $\vec{k}_1$. 
The subscripts on $r$ and $\delta$ indicate evaluating at the incident 
angle for wave \textcircled{$\scriptstyle 6$}, 
$\theta_{i,6}=90^{\circ}-\alpha$.
This allows us to write the reflected and evanescent field amplitudes in
terms of $A_6$,
\begin{equation}
\label{A1A6}
A_1 = A_6 e^{i\delta_6} e^{- i k_{6y} \frac{a}{\sqrt{3}}}, \quad
F_6 = A_6 (1+e^{i\delta_6}) e^{-i k_{6y} \frac{a}{2\sqrt{3}}} .
\end{equation}
%

An equivalent algebra applies for analyzing how wave 
\textcircled{$\scriptstyle 5$} incident on $b_0$ produces reflected wave 
\textcircled{$\scriptstyle 2$}, and an exterior evanescent wave 
\textcircled{$\scriptstyle 5'$}, in terms of incident angle 
$\theta_{i,5}=\alpha-150^{\circ}$, and associated phase shift, $\delta_5$.
The reflected and exterior field amplitudes are
\begin{equation}
\label{A2A5}
A_2 = A_5 e^{i\delta_5} e^{- i k_{5y} \frac{a}{\sqrt{3}}}, \quad
F_5 = A_5 (1+e^{i\delta_5}) e^{-i k_{5y} \frac{a}{2\sqrt{3}}} .
\end{equation}
Finally, wave \textcircled{$\scriptstyle 3$} incident on $b_0$ produces 
reflected wave \textcircled{$\scriptstyle 4$} and evanescent wave 
\textcircled{$\scriptstyle 3'$}, with similar expressions,
\begin{equation}
\label{A4A3}
A_4 = A_3 e^{i\delta_3} e^{- i k_{3y} \frac{a}{\sqrt{3}}}, \quad
F_3 = A_3 (1+e^{i\delta_3}) e^{-i k_{3y} \frac{a}{2\sqrt{3}}} .
\end{equation}
%

The same type of analysis can be applied to the waves incident on boundaries
$b_1$ and $b_2$, using the incident angles seen in Fig.\ \ref{reflections}.
One can also use the symmetries under $120^{\circ}$ rotations $R$, seeing that,
for example, $R \cdot b_0 = b_1$, and $R \cdot \vec{k}_6 = \vec{k}_4$, 
$R \cdot \vec{k}_1=\vec{k}_5$, implying that the phase relationship between 
waves \textcircled{$\scriptstyle 6$} and \textcircled{$\scriptstyle 1$} on 
$b_0$ is the same as the relationship between \textcircled{$\scriptstyle 4$}
and \textcircled{$\scriptstyle 5$} on $b_2$.
Similar arguments apply to the other pairs of waves.
The results of this analysis are summarized in Table \ref{matching}, where
the phase factors relating the wave amplitudes are denoted by
\begin{equation}
\label{Delta}
\Delta_l \equiv \delta_l - k_{ly}\frac{a}{\sqrt{3}}, \qquad l=3, 5, 6.
\end{equation}
There are only three distinct incident angles in the problem, 
corresponding to the three fundamental $\theta_i$ of Eq.\ \ref{inc356}.
These factors $\Delta_3$, $\Delta_5$ and $\Delta_6$ are now the only
quantities needed to correctly match self-consistently the interior
fields, according to the nine equations in the last column of Table
\ref{matching}.

\section{Determination of the resonant frequencies and wavefunctions}
\label{solve-omega}
The nine equations in the Table \ref{matching} can be summarized as
three basic ratios,
\begin{subequations}
\label{rats}
\begin{equation}
\frac{A_4}{A_3} = \frac{A_2}{A_1} = \frac{A_6}{A_5} = e^{i\Delta_3},
\end{equation}
\begin{equation}
\frac{A_2}{A_5} = \frac{A_6}{A_3} = \frac{A_4}{A_1} = e^{i\Delta_5},
\end{equation}
\begin{equation}
\frac{A_1}{A_6} = \frac{A_5}{A_4} = \frac{A_3}{A_2} = e^{i\Delta_6},
\end{equation}
\end{subequations}
Due to the simple structure, one sees the relation
\begin{equation}
\frac{A_2 A_4 A_6}{A_1 A_3 A_5} = e^{i3\Delta_3}=e^{i3\Delta_5}=e^{-i3\Delta_6}.
\end{equation}
This leads to two fundamental relations for these phases,
\begin{equation}
\label{phase35}
e^{i3(\Delta_3+\Delta_6)} = 1,  \qquad
e^{i3(\Delta_5+\Delta_6)} = 1.
\end{equation}
Using all of the equations (\ref{rats}) together gives the relation,
\begin{equation}
\label{phase356}
e^{i(\Delta_3+\Delta_5+2\Delta_6)} = 1.
\end{equation}
Although these last three relations are not linearly independent, all
are required to describe the solution.
The first two imply introduction of some integers denoted as $n_3$
and $n_5$, such that
\begin{equation}
\label{n35}
\Delta_3 + \Delta_6 = \frac{2\pi}{3} n_3,  \qquad
\Delta_5 + \Delta_6 = \frac{2\pi}{3} n_5.
\end{equation}
Eq.\  (\ref{phase356}) leads to,  on the other hand, 
\begin{equation}
\label{n6}
\Delta_3 + \Delta_5 + 2\Delta_6 = 2\pi n_6,
\end{equation}
where $n_6$ must also be an integer.
Comparing these equations demonstrates the constraint,
\begin{equation}
\label{constraint}
n_3 + n_5 =  3 n_6,
\end{equation}
that is, the sum of $n_3$ and $n_5$ must be a multiple of 3.
However, not all possible choices of these integers will lead to allowed
solutions.
Once the allowed quantum numbers $n_3$ and $n_5$ are determined,
the amplitude ratios of the six plane waves will be determined.

Eqs. (\ref{n35}) and (\ref{n6}) merely give some sums of the $\Delta_l$
phase factors, whereas, we actually need to determine each one separately,
and more importantly, we need to find the wavevector magnitude $k$.
This is accomplished by using their definitions (\ref{Delta}) together with
the corresponding $y$-components of the wavevectors, obtained from (\ref{kcomps}).
The necessary components are
\begin{subequations}
\label{y-comps}
\begin{equation}
k_{3y} = k\sin(\alpha-120^{\circ}) = -\frac{k}{2}(\sin\alpha+\sqrt{3}\cos\alpha),
\end{equation}
\begin{equation}
k_{5y} = k\sin(\alpha+120^{\circ}) = -\frac{k}{2}(\sin\alpha-\sqrt{3}\cos\alpha),
\end{equation}
\begin{equation}
k_{6y} = k\sin(-\alpha) = -k\sin\alpha.
\end{equation}
\end{subequations}
Then this leads to the $\Delta_l$ combinations,
\begin{subequations}
\begin{equation}
\label{Delta36}
\Delta_3+\Delta_6 = (\delta_3+\delta_6)
                  + \frac{1}{2}ka(\sqrt{3}\sin\alpha+\cos\alpha),
\end{equation}
\begin{equation}
\label{Delta56}
\Delta_5+\Delta_6 = (\delta_5+\delta_6)
                  + \frac{1}{2}ka(\sqrt{3}\sin\alpha-\cos\alpha),
\end{equation}
\begin{equation}
\label{Delta356}
\Delta_3+\Delta_5+2\Delta_6 = (\delta_3+\delta_5+2\delta_6)
                  + \sqrt{3} ~ ka \sin\alpha .
\end{equation}
\end{subequations}
The last of these then determines $k_y\equiv k_{1y}=k\sin\alpha$. 
Another combination involves only $k_x\equiv k_{1x}=k\cos\alpha$,
\begin{equation}
\label{Delta35}
\Delta_3-\Delta_5 = (\delta_3-\delta_5) + ka\cos\alpha = \frac{2\pi}{3}(n_3-n_5).
\end{equation}
Then the basic wavevector components of wave \textcircled{$\scriptstyle 1$}
are expressed as
\begin{subequations}
\label{kxky}
\begin{equation}
k_x a = ka\cos\alpha = \frac{2\pi}{3}(n_3-n_5)-(\delta_3-\delta_5),
\end{equation}
\begin{equation}
k_y a = ka\sin\alpha = \frac{1}{\sqrt{3}} \left[ 
 2\pi n_6 -(\delta_3+\delta_5+2\delta_6) \right].
\end{equation}
\end{subequations}
Remembering that the Fresnel phase shifts $\delta(\theta_{i,l})$ depend
ultimately on $\alpha$, via equations (\ref{inc356}), the Eqs.\ (\ref{kxky})
are seen to be coupled transcendental equations for unknowns $k$ and $\alpha$,
assuming $n_3$ and $n_5$ are given.
They can be solved in various ways.
A simple approach is to eliminate $k$, and then determine the allowed $\alpha$
as the roots of the following function,
\begin{eqnarray}
\label{p-alpha}
p(\alpha) &=& 
\left[\frac{2\pi}{3}(n_3+n_5)-(\delta_3+\delta_5+2\delta_6)\right]\cos\alpha
\\
&-&\sqrt{3}\left[\frac{2\pi}{3}(n_3-n_5)-(\delta_3-\delta_5)\right]\sin\alpha = 0.
\nonumber
\end{eqnarray}
In this last expression $n_6$ was eliminated using the constraint 
(\ref{constraint}).
In the general case, it is not possible to solve for $\alpha$ in closed form.
On the other hand, some straightforward analysis of this function, together
with numerical evaluation indicates the region in which to look for the
quantum numbers $n_3$ and $n_5$.
%

Once $\alpha$ has been determined, the modulus of the mode's wavevector, $k$,
can be found from either (\ref{kxky}a) or (\ref{kxky}b), or by their combination,
written in a form like that for the known solution for the DBC case 
[Eq.\ (\ref{kDBC})],
\begin{eqnarray}
\label{wavevec}
ka=\frac{2\pi}{3} &\Big\{ &
  \Big[ n_3-n_5-\frac{3}{2\pi}(\delta_3-\delta_5) \Big]^2 \nonumber \\
&&+3\Big[n_6-\frac{1}{2\pi}(\delta_3+\delta_5+2\delta_6) \Big]^2 \Big\}^{1/2}.
\end{eqnarray}
%

\subsection{The resonant wavefunctions}
Assuming $\alpha$ has been found (see below for DBC, TE and TM cases), then
$k_x$, $k_y$, and $k$ or $\omega=c^{*} k$ are determined.
In addition, Eqs.\ (\ref{Delta}) determine the phase factors needed to
express the complete wavefunction for any mode, in terms of $\alpha$ via
Eqs.\ (\ref{kxky}) and (\ref{inc356}).
The quantum numbers and Fresnel phase shifts give the $\Delta_l$ as
\begin{subequations}
\begin{equation}
\Delta_3=\frac{1}{3} \left[ \frac{2\pi}{3}(2n_3-n_5)
                           +(\delta_3+\delta_5-\delta_6) \right],
\end{equation}
\begin{equation}
\Delta_5=\frac{1}{3} \left[ \frac{2\pi}{3}(2n_5-n_3)
                           +(\delta_3+\delta_5-\delta_6) \right],
\end{equation}
\begin{equation}
\Delta_6=\frac{1}{3} \left[ \frac{2\pi}{3}(n_3+n_5)
                           -(\delta_3+\delta_5-\delta_6) \right],
\end{equation}
\end{subequations}
all of which depend on a net phase factor,
\begin{equation}
\varphi(\alpha) \equiv \delta_3+\delta_5-\delta_6.
\end{equation}
This allows evaluation of the wavefunction (\ref{psi-six}),
simplified by freely choosing the amplitude of wave 
\textcircled{$\scriptstyle 1$} as
\begin{equation}
\label{freely}
A_1 = \frac{1}{2} A_0 \exp\left\{ -i\left[ 
       \frac{2\pi}{9}(n_3+n_5)+\frac{1}{6}\varphi \right] \right\},
\end{equation}
where $A_0$ is the overall wavefunction amplitude.
\begin{eqnarray}
\label{psi}
&\psi& = A_0 \left\{ 
    e^{ik_{x}x}  \cos\left[k_{y}y -\frac{2\pi}{9}(n_3+n_5)
                                -\frac{1}{6}\varphi \right]
\right.  \nonumber \\
&+& e^{ik_{3x}x} \cos\left[k_{3y}y-\frac{2\pi}{9}(n_5-2n_3)
                                 -\frac{1}{6}\varphi  \right]
\nonumber \\
&+& 
\left.
    e^{ik_{5x}x} \cos\left[k_{5y}y-\frac{2\pi}{9}(n_3-2n_5)
                                  -\frac{1}{6}\varphi \right]
\right\},
\end{eqnarray}
where the wavevectors of waves \textcircled{$\scriptstyle 1$},
\textcircled{$\scriptstyle 3$}, and \textcircled{$\scriptstyle 5$} are
needed, e.g., from Eqs.\ (\ref{kxky}) and using Table \ref{6waves},
\begin{equation}
k_{3x} = -\frac{1}{2} k_x + \frac{\sqrt{3}}{2} k_y, \qquad
k_{3y} = -\frac{\sqrt{3}}{2} k_x - \frac{1}{2} k_y.
\end{equation}
\begin{equation}
k_{5x} = -\frac{1}{2} k_x - \frac{\sqrt{3}}{2} k_y, \qquad
k_{5y} =  \frac{\sqrt{3}}{2} k_x - \frac{1}{2} k_y,
\end{equation}
%

Note that if $\alpha=60^{\circ}$, then
$\theta_{i,3}=\theta_{i,6}=30^{\circ}$, making $\delta_3=\delta_6$, and hence,
$\varphi=\delta_5 = -\pi$ (grazing incidence for wave
\textcircled{$\scriptstyle 5$}, TE or TM polarization).
Putting $\alpha=60^{\circ}$ also gives $n_3=2n_5+3$ (See \ref{DBC}
below).
These values cause the wavefunction $\psi$ to vanish;  there are no
modes with $\alpha=60^{\circ}$ for the DBC, TE or TM cases, although
it is tempting to sketch such ray diagrams.

\subsection{Equilateral triangle cavity with Dirichlet boundary conditions}
\label{DBC}
It is interesting to check the validity of the six-wave analysis in
the case of Dirichlet boundary conditions; this also helps to locate
the allowed $n_3$ and $n_5$ for Maxwell boundary conditions.
The net field on the boundary becomes zero when the Fresnel phase shifts
are all taken to be $\delta_l=-\pi$ (this is the limiting phase shift
for grazing incidence, in either TE or TM polarization).
Then the wavevector components reduce to
\begin{equation}
\label{kxkyDBC}
k_x a = \frac{2\pi}{3}(n_3-n_5), \qquad k_y a = \frac{2\pi}{\sqrt{3}}(n_6+2).
\end{equation}
The original assumption, $\alpha\ge 60^{\circ}$, imposes the relation
$\tan\alpha=k_y/k_x\ge\sqrt{3}$. 
At the limiting value $\alpha=60^{\circ}$, however, the net wavefunction 
$\psi$ of Eq.\ (\ref{psi}) vanishes; $\alpha=60^{\circ}$ is not allowed.
There results $n_3 < 2n_5+3$ (equality maps to $\alpha=60^{\circ}$).
Also it was implicitly assumed that $k_x\ge0$, hence, we require $n_3\ge n_5$.
Therefore, a given choice of $n_5$ allows a limited range of $n_3$,
\begin{equation}
\label{limited}
n_5 \le n_3 < 2n_5+3.
\end{equation}
Using values $n_5\ge0$ and evaluating the possible quantum numbers, the 
well-known solutions for DBC are recovered\cite{Lame52,BB97,Chang00}, 
in terms of shifted quantum numbers, 
\begin{equation}
\label{mndef}
m\equiv(n_3-n_5), \qquad n\equiv(n_6+2).
\end{equation}
These must be both odd, or both even, with the restriction 
$m<n$.
Generally, we use these quantum numbers to label the
modes with dielectric boundary conditions, in place of $n_3, n_5$.
Then the DBC wavevector moduli are reproduced from (\ref{wavevec}),
\begin{equation}
\label{kDBC}
ka=\frac{2\pi}{3} \{ m^2 +3n^2 \}^{1/2}.
\end{equation}
Note that the prohibited solutions at $\alpha=60^{\circ}$ would correspond
to the prohibited case, $m=n$.
It is straightforward to check that the ($m<n$) DBC wavefunctions vanish
on all the boundaries.
%

\subsection{Maxwell boundary conditions: TM or TE polarization}
To define Eq.\ (\ref{p-alpha}) for $p(\alpha)=0$,
we require the Fresnel phase shifts for MBC.
For TM polarization, $\psi=E_z$, and the Fresnel reflection coefficient for 
electric field polarized perpendicular to the plane of incidence is needed.
For incident angle $\theta_i$, the formula is\cite{Jackson1}
\begin{equation}
\label{FresnelTM}
\frac{\psi_{\mathrm{refl.}}}{\psi_{\mathrm{inc.}}} =
e^{i \delta} = \frac{\sqrt{\frac{\epsilon}{\mu}}\cos\theta_i 
                     -\sqrt{\frac{\epsilon'}{\mu'}}\cos\theta'}
                     {\sqrt{\frac{\epsilon}{\mu}}\cos\theta_i 
                     +\sqrt{\frac{\epsilon'}{\mu'}}\cos\theta'}
\qquad \mathrm{(TM)}.
\end{equation}
Unprimed electric permittivity $\epsilon$ and magnetic permeability $\mu$
correspond to the cavity medium, whereas the primed values are those outside
the cavity; the refractive indexes result from $\mathsf{n}=\sqrt{\epsilon\mu}$.
Angle $\theta'$ is the refraction angle, a complex quantity obtained from
Snell's Law under TIR,
\begin{equation}
\label{gamma}
\cos\theta' = i\gamma' \equiv i 
\sqrt{\left(\sin\theta_i / \sin\theta_c \right)^2-1},
\end{equation}
where critical angle $\theta_c$ is defined in Eq.\ (\ref{crit}).
The phase shift can be expressed also via
\begin{equation}
\label{deltaTM}
\tan\frac{\delta}{2} = -\frac{\mu}{\mu'} 
                       \sqrt{\frac{\cos^{2}\theta_c}{\cos^{2}\theta_i}-1}
\qquad \mathrm{(TM)}.
\end{equation}
In the usual case for many optical materials with $\mu\approx\mu'\approx 1$,
the phase shift changes slowly as $\theta_i$ ranges from $\theta_c$ ($\delta=0$)
to $90^{\circ}$ ($\delta=-\pi$).
Typical examples for index ratio $\mathsf{N}=2.5, 4.0, 10.0$ are shown in Fig.\
\ref{delta}.
In numerical results presented here we assume $\mu=\mu'=1$, as for most 
optical materials.

For TE polarization, $\psi=H_z$, and the Fresnel reflection coefficient 
for electric field polarized within the plane of incidence is needed.
The Fresnel formula is\cite{Jackson1}
\begin{equation}
\label{FresnelTE}
\frac{\psi_{\mathrm{refl.}}}{\psi_{\mathrm{inc.}}} = e^{i\delta} =
 \frac{\sqrt{\frac{\epsilon'}{\mu'}}\cos\theta_i 
 -\sqrt{\frac{\epsilon}{\mu}}\cos\theta'}
 {\sqrt{\frac{\epsilon'}{\mu'}}\cos\theta_i 
 +\sqrt{\frac{\epsilon}{\mu}}\cos\theta'}
\qquad \mathrm{(TE)}.
\end{equation}
Equivalently, the reflection phase shift is given by
\begin{equation}
\label{deltaTE}
\tan\frac{\delta}{2} = -\frac{\epsilon}{\epsilon'} 
                       \sqrt{\frac{\cos^{2}\theta_c}{\cos^{2}\theta_i}-1}
\qquad \mathrm{(TE)}.
\end{equation}
The important difference, in comparison with that for TM polarization, is
the presence of the factor $\epsilon/\epsilon'$ instead of $\mu/\mu'$.
This significantly enhances the initial rate at which $\delta$ increases with
$\theta_i$, as can be seen in Fig.\ \ref{delta}.
Both the TM and TE phases shifts reach $-\pi$ at $\theta_i=90^{\circ}$, but for 
TE the approach is much more rapid.
This can also be taken to imply that a DBC approximation for the modes, such as 
that applied in Ref.\ \onlinecite{Wysin05a}, is more reasonable for
TE modes than for TM modes.

\begin{figure}
\includegraphics[angle=-90.0,width=\columnwidth]{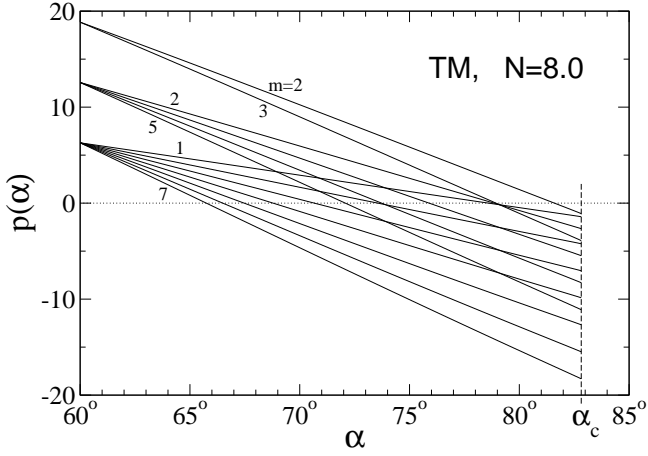}
\caption{ \label{palpha8TM} Example $p(\alpha)$ functions [Eq.\ (\ref{p-alpha})]
at $\mathsf{N}=8$, $\mu=\mu'$, for TM polarization, 
giving the lowest 13 valid mode solutions,  located
by zero crossings (dotted line).  The individual curves are labeled by $(m,n)$, 
see Eq.\ (\ref{mndef}), where $m$ changes in unit increments.  
There are three different groupings of curves.  The bottom group (7 curves) 
has $n-m=2$, the middle group has $n-m=4$, and the top group has $n-m=6$.
The values at the left limit are $p(60^{\circ})=(n-m)\pi$. The right limiting
point is $\alpha_c=90^{\circ}-\theta_c$.
}
\end{figure}

\begin{figure}
\includegraphics[angle=-90.0,width=\columnwidth]{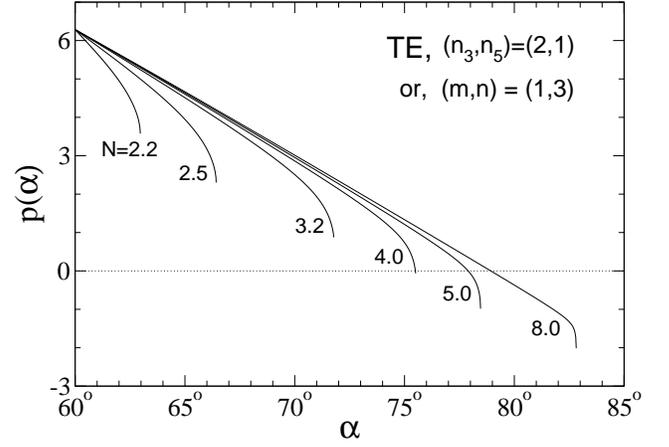}
\caption{ \label{palphaTE} Example $p(\alpha)$ functions for TE polarization, 
with the quantum numbers fixed at their lowest nonzero values, for various
indicated index ratios $\mathsf{N}$.  Only the curves which cross zero 
(dotted line) give valid mode solutions ($\mathsf{N}\agt 4$). The different 
termination points of the curves occur at $\alpha_c=90^{\circ}-\theta_c$.
}
\end{figure}

\subsection{Mode solutions: determination of $\alpha$ and $(n_3,n_5)$ pairs,
Maxwell boundary conditions}
Now consider the solution of Eq.\ (\ref{p-alpha}) for $\alpha$ under
MBC.
As shown earlier, there are no solutions with $\alpha=60^{\circ}$, because
such a choice causes the wavefunction (\ref{psi}) to vanish.
The other extreme, $\alpha=90^{\circ}$, which requires $n_3=n_5$, is 
prohibited because wave \textcircled{$\scriptstyle 6$} can never experience 
TIR at a vanishing incident angle.
Then, for MBC, similar to Eq.\ (\ref{limited}),
the search for possible $(n_3,n_5)$ pairs must take place in the range
\begin{equation}
\label{dielimited}
n_5 < n_3 < 2n_5+3.
\end{equation}

\begin{figure}
\includegraphics[angle=-90.0,width=\columnwidth]{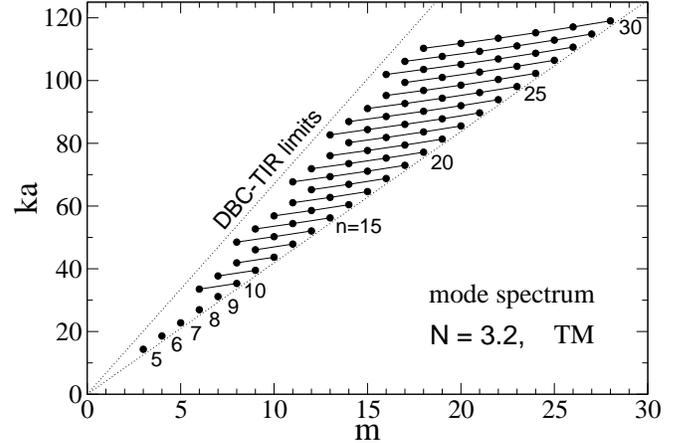}
\caption{ \label{TM3.2_ka} The lowest mode wavevectors for TM polarization
at index ratio $\mathsf{N}=3.2$, plotted versus quantum index $m=n_3-n_5$,
for different values of the other index $n=n_6+2$ (Eq.\ \ref{mndef})
indicated next to the curves, changing by unit increments.  The solid
circles indicate allowed modes; the solid lines connect those having
equal values of $n$.  The dotted lines locate the limits of stable 
TIR as expected from using the DBC solution for the ETR, 
Eqs.\ (\ref{l-limit}) and (\ref{r-limit}), explained in the text.
}
\end{figure}

It is straightforward to calculate $p(\alpha)$ [Eq.\ (\ref{p-alpha})]
numerically and determine the zero crossings, which only take place
provided that $n_3$ is fairly close to the upper limit of (\ref{dielimited}).
Choices of $n_3, n_5$ were made as follows.
Starting from some $n_5\ge 0$, calculate $n_3=2n_5+3$, which gives a
prohibited pair at $\alpha=60^{\circ}$.
Automatically the sum $(n_3+n_5)$ is a multiple of 3.
Then the first pair to check for a valid solution, is to take $n_3$ reduced by 1
($n_3\to n_3-1$) and $n_5$ increased by 1 ($n_5\to n_5+1$), such that the sum is 
the same multiple of 3.
The resulting pair $(n_3,n_5)$ will likely have a solution for $\alpha$ if
$n_5$ is adequately large.
This initial pair is equivalently set by
\begin{equation}
n_3=2n_5, \qquad n_5 = 1, 2, 3 \ldots
\end{equation}
Other pairs to try are found by continuing the reduction of $n_3$ by 1
together with the incrementing of $n_5$ by 1.
%

The solution for $\alpha$ must occur within the range
\begin{equation}
60^{\circ} < \alpha < 90^{\circ}-\theta_c, 
\end{equation}
because the incident angle of wave \textcircled{$\scriptstyle 6$}, 
$\theta_{i,6}=90^{\circ}-\alpha$, cannot surpass the critical angle.
Therefore, there is a considerably larger range for possible
solutions for $\alpha$ as the index ratio increases (smaller $\theta_c$).
Conversely, an index ratio only slightly above 2.0 makes a substantially
limited search range for $\alpha$, as a result, fairly large values of $n_5$
are required before a solution is found.

Plots of $p(\alpha)$ are surprisingly close to linear, especially for
larger index ratio $\mathsf{N}$ and TM polarization. 
Examples are shown in Fig.\ \ref{palpha8TM} for index ratio
$\mathsf{N}=8$, with various $(n_3,n_5)$ pairs [or in terms of $(m,n)$]
and TM polarization.
One should note that the various $p(\alpha)$ curves at different quantum
numbers are closely related; the obvious visible change occurs in the
typical slopes as $n_3$ and $n_5$ are adjusted, while the $\delta_i(\alpha)$
do not change.
The dependence of some $p(\alpha)$ curves on $\mathsf{N}$ is indicated
in Fig.\ \ref{palphaTE}, at fixed $(n_3,n_5)=(2,1)$ for TE polarization.
The terminating points of these curves occur at 
$\alpha_c=90^{\circ}-\theta_c$.
Then, the possibility for a zero crossing is greatly enhanced as
$\mathsf{N}$ increases.
Similar plots for TM polarization show curves much closer to linear form.
The strong downward curvature as $\alpha\to\alpha_c$ for TE
polarization means that for given quantum numbers, a TE polarized mode
can be excited at lower index ratio than the same TM mode.

\begin{figure}
\includegraphics[angle=-90.0,width=\columnwidth]{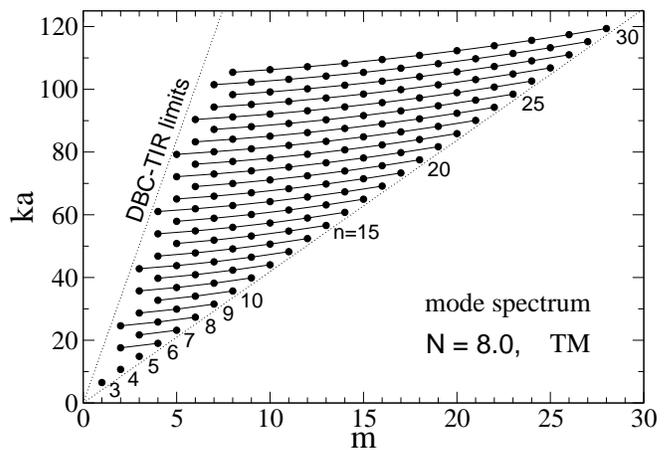}
\caption{ \label{TM8.0_ka} The lowest mode wavevectors for TM polarization
at index ratio $\mathsf{N}=8.0$, plotted as described in Fig.\ \ref{TM3.2_ka}.
Note also that the fundamental mode here has lower $(m,n)$ and $ka$ than that
for $\mathsf{N}=3.2$ .
}
\end{figure}

\section{Calculations of Mode spectra and properties}
Having found $(n_3,n_5)$ pairs and associated $\alpha$, we can look at 
the mode dependence on polarization and refractive index ratio $\mathsf{N}$.
Henceforth, modes will be labeled by the quantum number pairs, $(m,n)$.

\begin{figure}
\includegraphics[angle=0.0,width=\smallfigwidth]{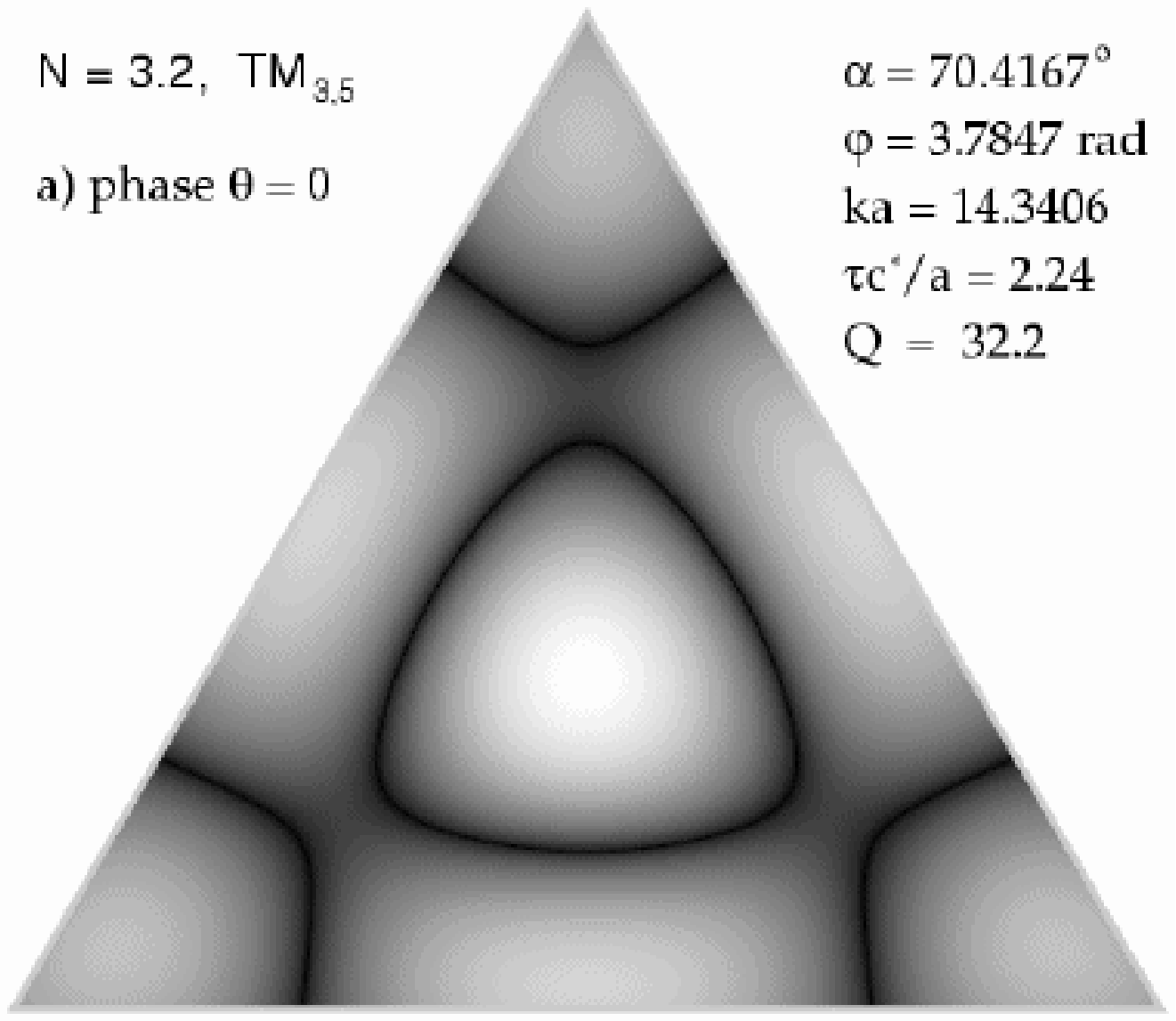}
\includegraphics[angle=0.0,width=\smallfigwidth]{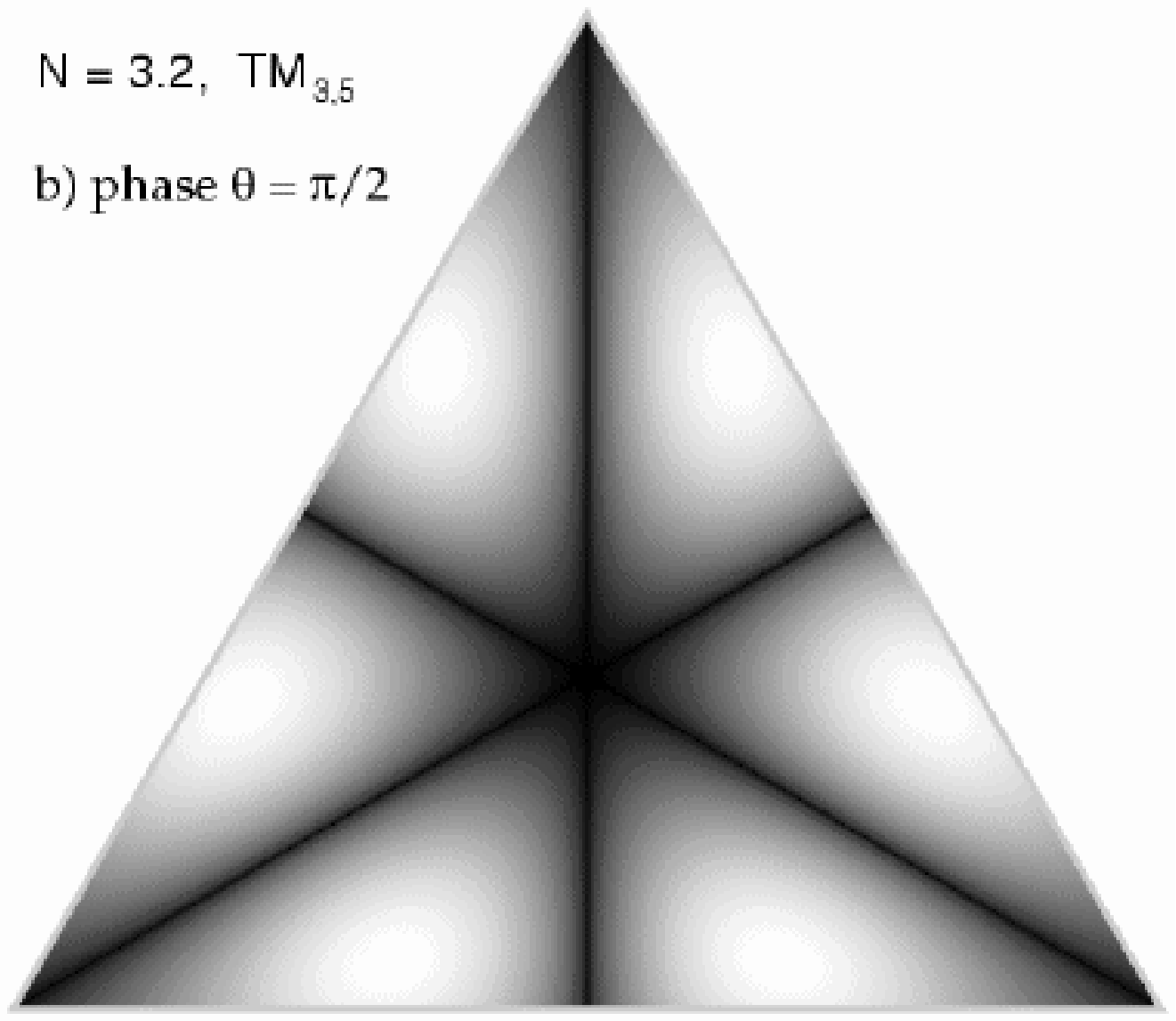}
\caption{ \label{TM32_m1}  Fundamental TM modes of oscillation at 
$\mathsf{N}=3.2$, with $(m,n)=(3,5)$, where the pixel intensity is 
proportional to $|Re\{\psi\}|^{1/2}$, which enhances the definition of the 
nodal curves.  The black nodal curves separate alternating regions of 
positive and negative  $Re\{\psi\}$.  Two degenerate wavefunctions are 
displayed.  In a) phase is $\theta_0=0$, in b) the phase is $\theta_0=\pi/2$, 
where the mode amplitude is $A_0=e^{i\theta_0}$, see text.  
}
\end{figure}

\begin{figure}
\includegraphics[angle=0.0,width=\smallfigwidth]{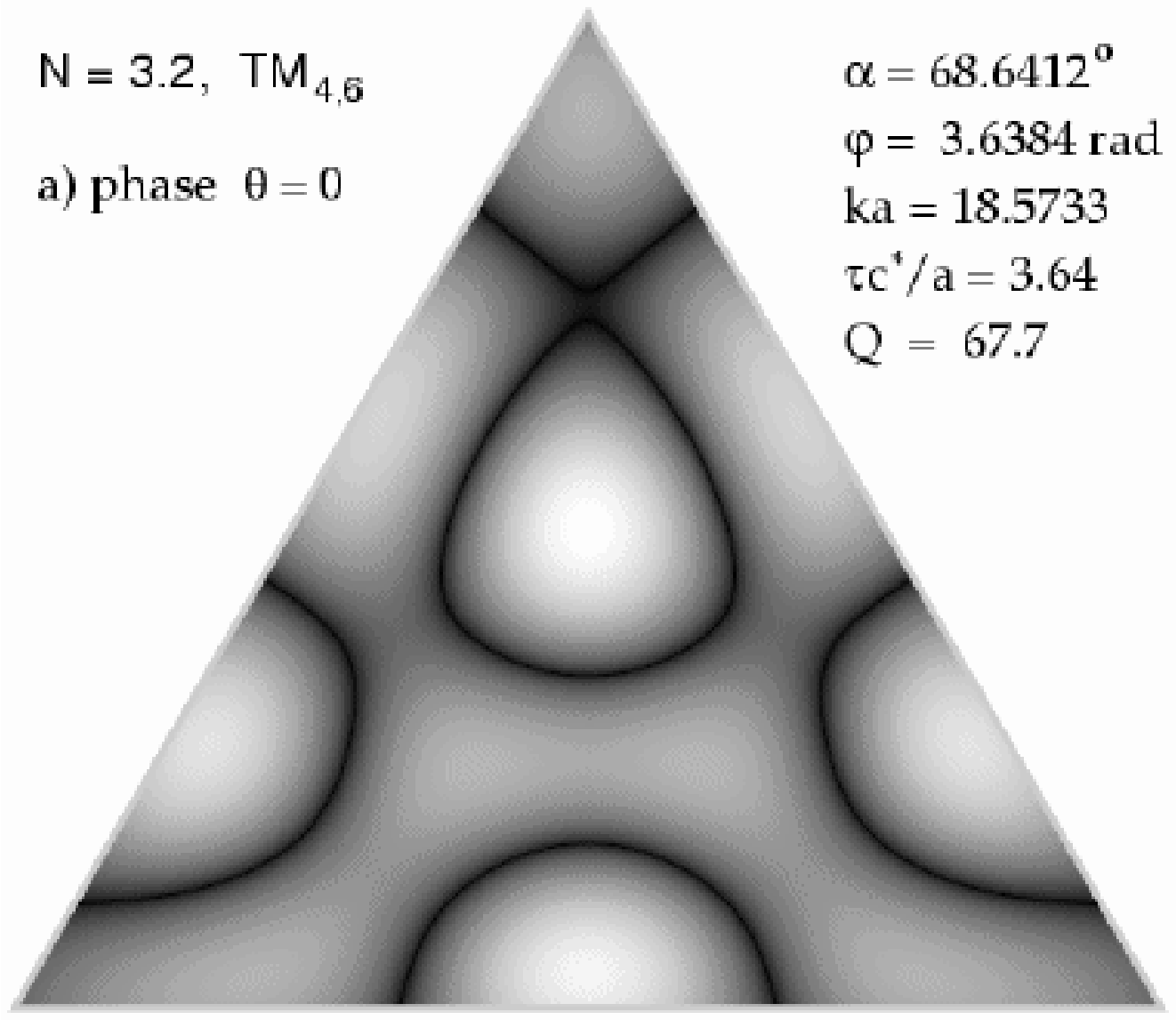}
\includegraphics[angle=0.0,width=\smallfigwidth]{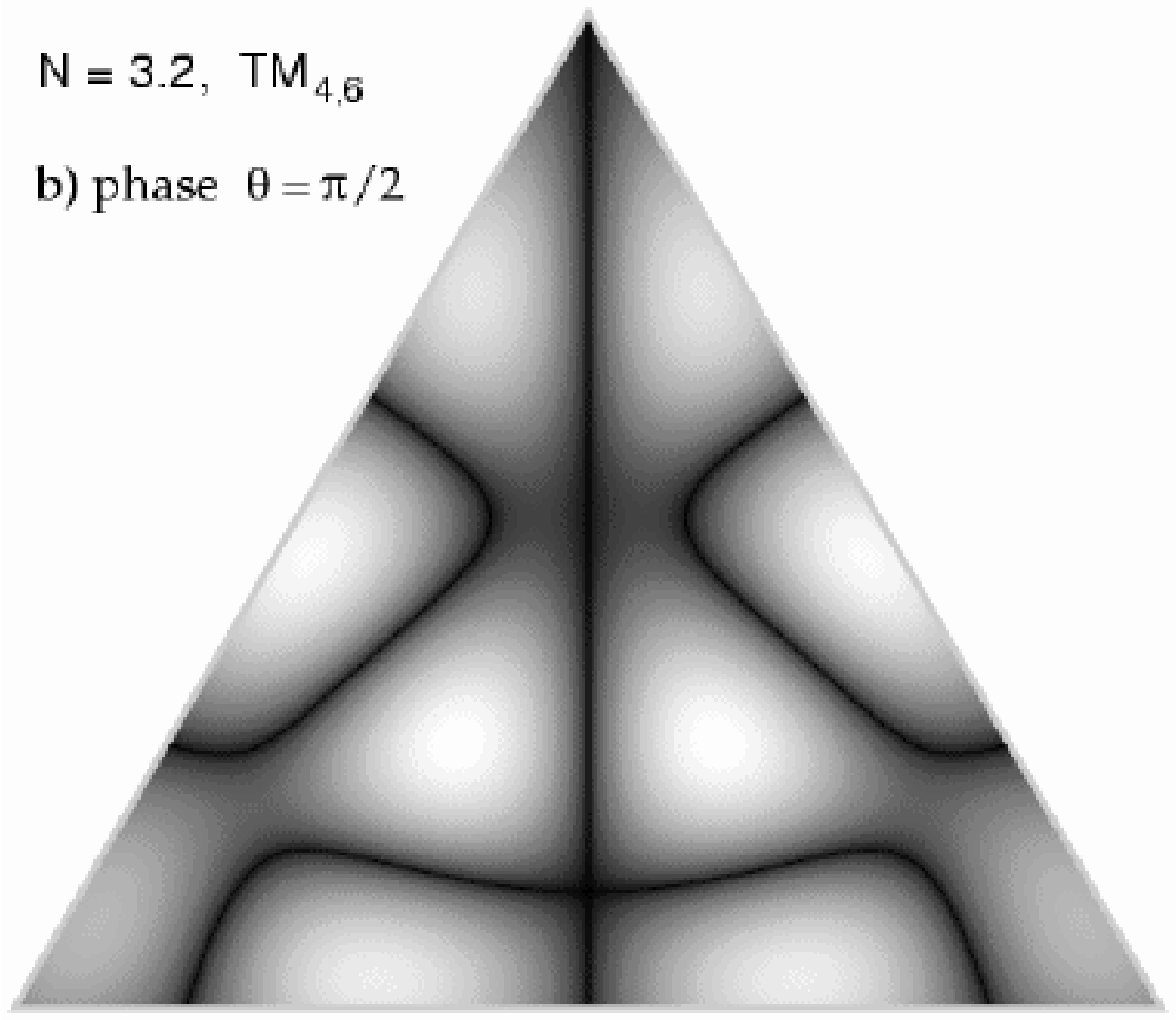}
\caption{ \label{TM32_m2}  First excited TM modes of oscillation 
at $\mathsf{N}=3.2$, with $(m,n)=(4,6)$, as described in Fig.\ \ref{TM32_m1}.
Two degenerate wavefunctions are displayed. In a) phase is $\theta_0=0$, 
in b) the phase is $\theta_0=\pi/2$, where the mode amplitude is 
$A_0=e^{i\theta_0}$, see text.  
}
\end{figure}

\subsection{TM polarization}
For TM polarization, the frequencies of the lowest modes are displayed 
in Fig.\ \ref{TM3.2_ka} for $\mathsf{N}=3.2$, in terms of their
dependences on the mode indexes $(m,n)$, defined in Eq.\ (\ref{mndef}).
This value for index ratio was used in Ref.\ \onlinecite{Huang+01} in
numerical and experimental studies of ETR semiconductor cavities,
with a different theoretical analysis of the modes.
In general, the trend is for $ka$ to increase with increasing 
quantum numbers.
Just as in the DBC analysis, the allowed modes for MBC must have
indexes $m$ and $n$ either both odd, or both even (parity constraint).
Note, however, that the TM mode wavevectors for chosen $(m,n)$ are
noticeably lower than the values expected from the simplified DBC theory,
expression (\ref{kDBC}), see Figs.\ \ref{TM-ka-N} and \ref{TE-ka-N} below.
This can be attributed to the fact that the reflection phase shifts
[from (\ref{deltaTM})] due not tend towards $-\pi$ even at very large
index ratio $\mathsf{N}$.

The positions of the $ka$ upper limits  (see dotted lines in Fig.\ 
\ref{TM3.2_ka}) can be estimated quite accurately by using the 
TIR cutoff for a mode from the DBC theory\cite{Wysin05a}.
Using the DBC solutions, all six plane waves can maintain TIR at all 
the boundaries only when
\begin{equation}
\label{DBC-limit}
m > n \sqrt{\frac{3}{\mathsf{N}^2-1}}.
\end{equation}
Combining this with Eq.\ (\ref{kDBC}) for $ka$ leads to the DBC-TIR
limiting curve, expected to hold reasonably well at higher index ratio,
\begin{equation}
\label{l-limit}
ka = \frac{\omega a}{c^*} < \frac{2\pi \mathsf{N}}{3} m.
\end{equation}
Additionally, the right limiting points of each curve correspond to 
quantum index pairs with $n=m+2$, or equivalently, $n_3=2n_5$.
Then also using the DBC wavevectors (\ref{kDBC}), together with constraint
$m<n$, valid for either DBC or MBC, leads to the $ka$ lower limit,
\begin{equation}
\label{r-limit}
ka = \frac{\omega a}{c^*} > \frac{4\pi}{3} m.
\end{equation}
Indeed, all the results for Maxwell boundary conditions lie between
these results, the dotted lines in Fig.\ \ref{TM3.2_ka}, lending 
support to general aspects of the DBC theory.

\begin{figure}
\includegraphics[angle=-90.0,width=\columnwidth]{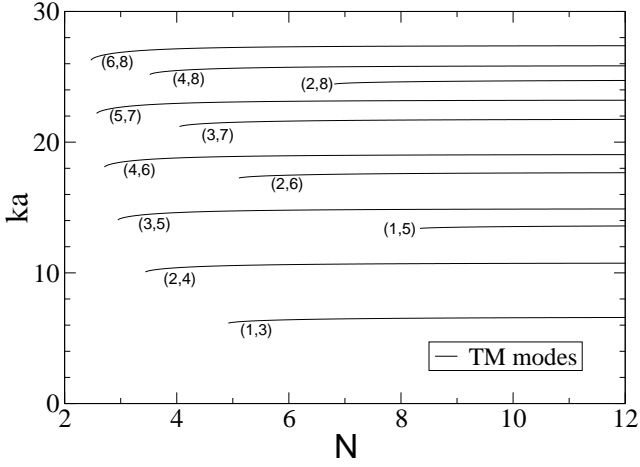}
\caption{ \label{TM-ka-N} TM mode wavevectors as functions of the
index ratio $\mathsf{N}$, for modes indicated by quantum indexes $(m,n)$.
}
\end{figure}

As $\mathsf{N}$ is increased, the lower TIR limit does not change,
while the DBC-TIR upper limit moves steeper, or to the upper left, 
encompassing more of the possible modes from the DBC theory, 
down towards smaller $m$ for a given $n$.
An example of this is given in Fig.\ \ref{TM8.0_ka}, showing the 
corresponding results for index ratio $\mathsf{N}=8$.
A larger number of states appears for each $n$, although
the mode wavevectors have changed slightly.\footnote{The mode frequencies,
given by $\omega=ck/\mathsf{n}$, may or may not diminish with increasing
$\mathsf{N}$, depending on whether $\mathsf{N}$ changed due to increased 
$\mathsf{n}$ or due to decreased $\mathsf{n'}$.}
Conversely, as $\mathsf{N}$ decreases below the value 2.0, the DBC-TIR
upper limit passes the lower limit, leaving no modes that can be confined 
by TIR.
Furthermore, as long as $\mathsf{N}>2.0$, the total number of modes
is not finite, since $n$ can be adjusted to an adequately large value
to reach the fundamental mode.
The main effect of placing $\mathsf{N}$ very close to $2.0$ will be to
force the lowest frequency mode to a large value of $ka$ and associated large
minimum value of $n$.

The fundamental mode quantum indexes depend on $\mathsf{N}$.
At $\mathsf{N}=3.2$, the fundamental mode has $(m,n)=(3,5)$,
and a diagram of its interior wavefunction ($\psi=E_z$) is shown in 
Fig.\ \ref{TM32_m1}.
The intensity of the pixels in these images has been set proportional
to $\vert \mathrm{Re}\{\psi\}\vert^{1/2}$, rather than linear in 
$\mathrm{Re}\{\psi\}$, in order to sharpen the appearance of the zero 
crossings.
The resulting nodal curves separate neighboring positive/negative
regions of the wavefunctions.
The calculation of mode lifetimes and quality factors indicated
on the wavefunction diagrams is described later in Sec.\ \ref{life}.

Wavefunctions for the first excited TM state at $\mathsf{N}=3.2$,
with $(m,n)=(4,6)$, are shown in Fig.\ \ref{TM32_m2}.
It is important to note that for the TM modes, the fields have substantial
nonzero amplitudes even at the cavity edges.
Furthermore, all these TIR-confined modes are doubly degenerate, since one
can choose the two values $\pm k_x$ and form degenerate pairs of states.
Alternatively, different degenerate wavefunctions can be obtained by 
putting the overall amplitude $A_0= e^{i\theta_0}$, choosing some 
arbitrary phase $\theta_0$, and using only the real part of (\ref{psi}).
The degenerate pairs presented here were formed using $\theta_0=0$ and
$\theta_0=\pi/2$.

%

\begin{figure}
\includegraphics[angle=-90.0,width=\columnwidth]{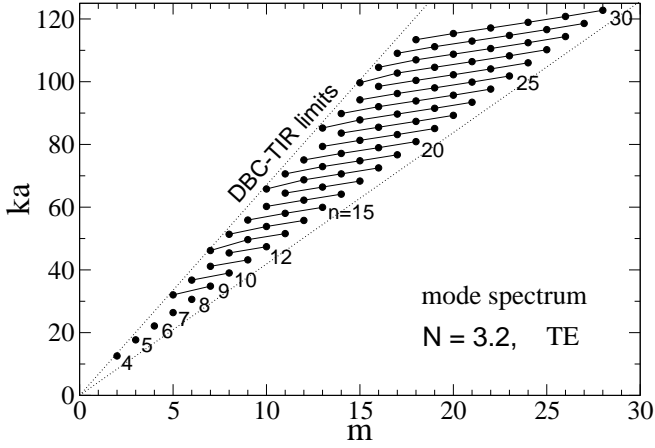}
\caption{ \label{TE3.2_ka} The lowest mode wavevectors for TE polarization
at index ratio $\mathsf{N}=3.2$, plotted as described in Fig.\ \ref{TM3.2_ka}.
The fundamental mode here has lower $(m,n)$ and $ka$ than that for TM 
polarization.  On the other hand, at fixed $(m,n)$, the mode wavevectors 
here are higher than those for TM polarization.
}
\end{figure}

\begin{figure}
\includegraphics[angle=0.0,width=\smallfigwidth]{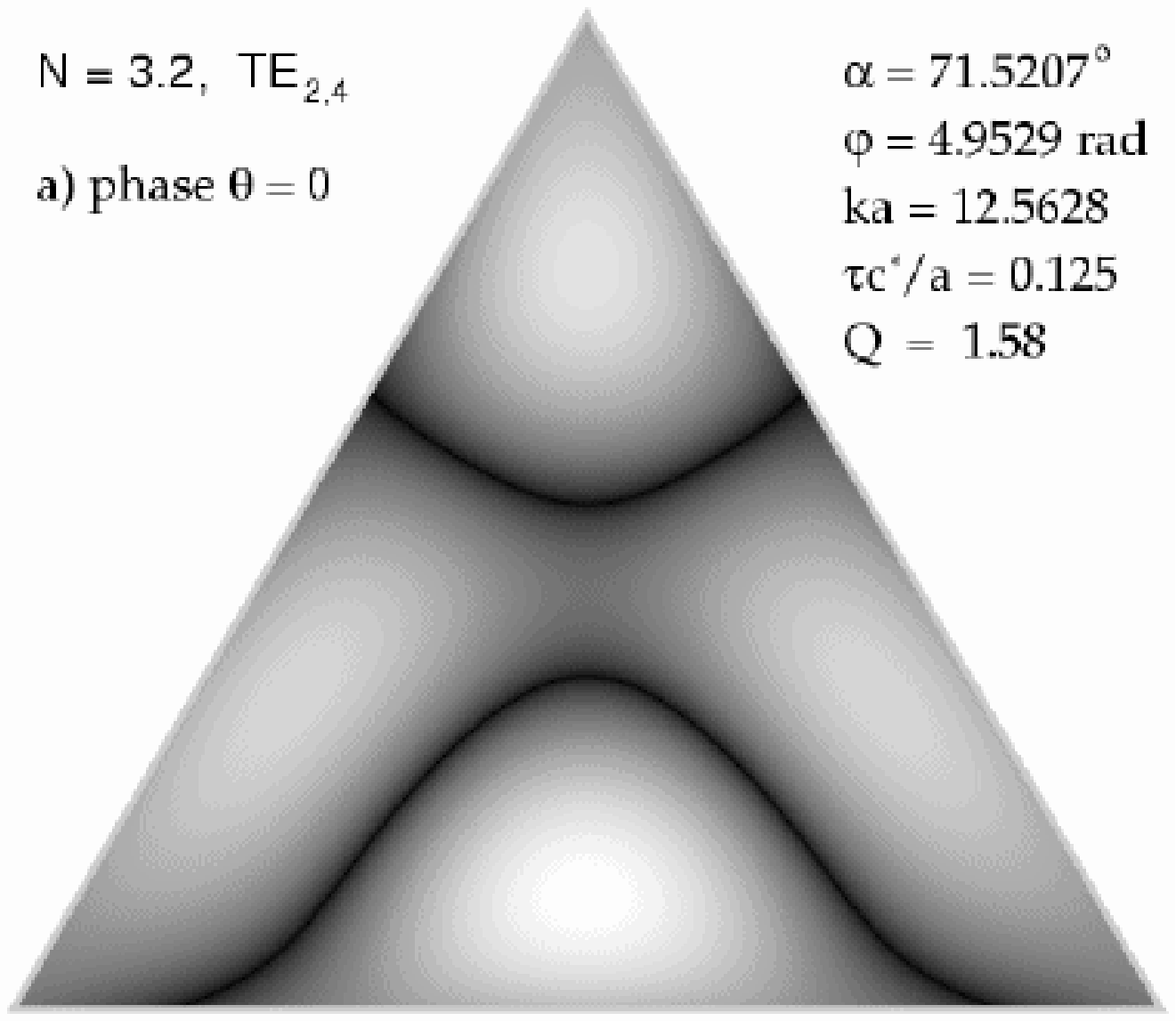}
\includegraphics[angle=0.0,width=\smallfigwidth]{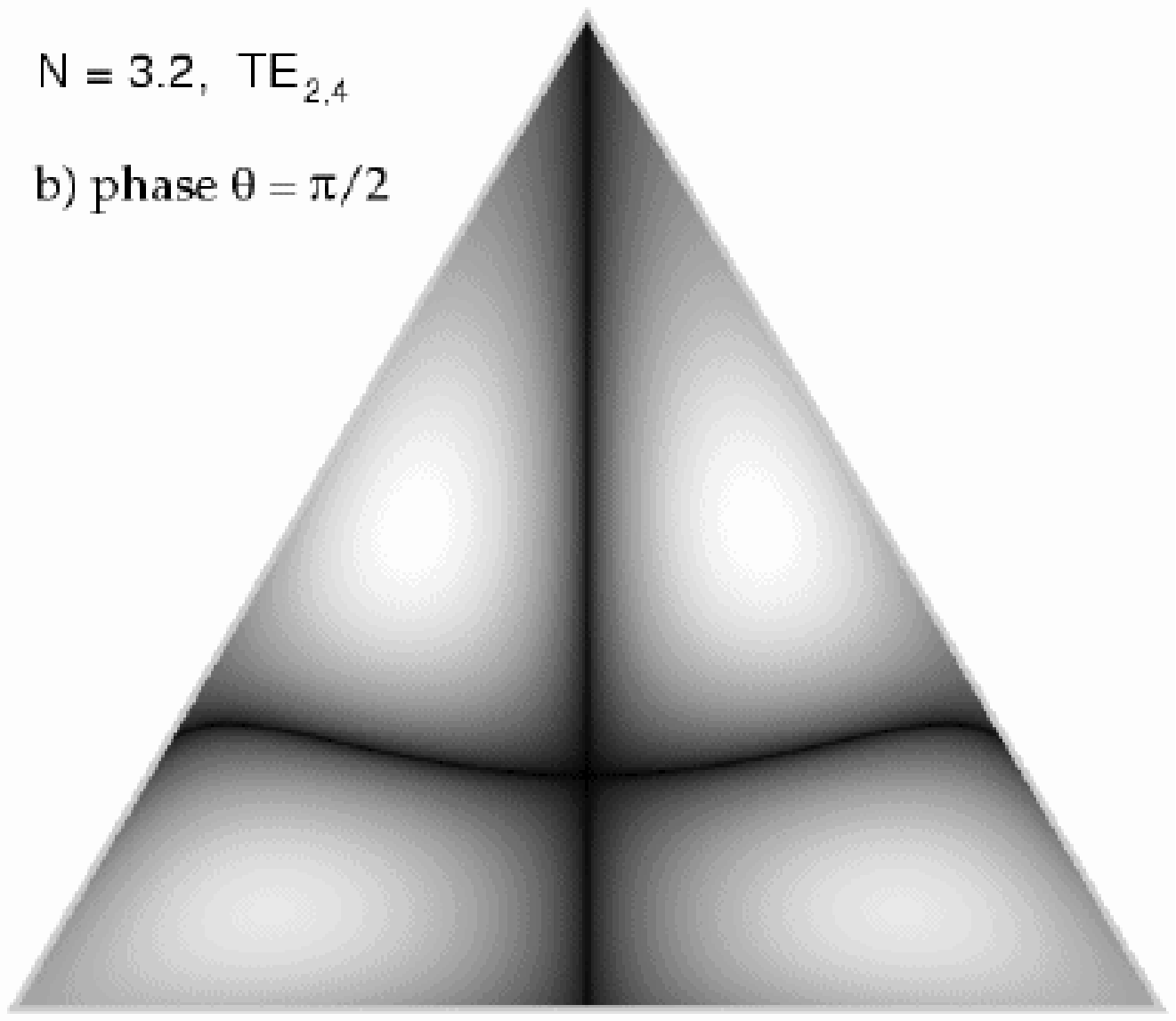}
\caption{ \label{TE32_m1} Fundamental TE modes of oscillation 
at $\mathsf{N}=3.2$, with $(m,n)=(2,4)$, as described in Fig.\ \ref{TM32_m1}.
Two degenerate wavefunctions are displayed. In a) phase is $\theta_0=0$, 
in b) the phase is $\theta_0=\pi/2$, where the mode amplitude is 
$A_0=e^{i\theta_0}$, see text.  
}
\end{figure}

A rather weak dependence of some of the TM$_{m,n}$  mode frequencies on 
index ratio is shown in Fig.\ \ref{TM-ka-N}.
The cut-off index ratios appear clearly as the left termination points
of each curve.
These cutoffs are similar in magnitude to that from the DBC theory, 
rewriting Eq.\ (\ref{DBC-limit}),
\begin{equation}
\label{DBC-cut}
\mathsf{N}>\mathsf{N}_c \equiv \sqrt{3\frac{n^2}{m^2}+1}.
\end{equation}
(See Fig.\ \ref{TE-ka-N} for the DBC cutoffs.)
However, the TM mode wavevectors are considerably lower than the prediction
of the DBC theory, Eq.\ (\ref{kDBC}), because the TM reflection phase shifts
$\delta_{i,l}$ are never near $-\pi$.

\subsection{TE polarization}
Results for the TE mode spectrum at $\mathsf{N}=3.2$ are shown in Fig.\ 
\ref{TE3.2_ka}, in the same manner as displayed for TM polarization.
Again, the allowed mode indexes satisfy the parity constraint, and the $ka$
values fall within the limits of the DBC-TIR theory.
On the other hand, there are two primary differences compared to TM 
polarization.
First, the fundamental mode has a lower $(m,n)$ pair, and in addition, 
a slightly lower $ka$ than for the fundamental mode with TM polarization.
Second, at a given $(m,n)$, we see that the wavevectors for TE polarization
always are noticeably higher than for TM polarization.
%

At $\mathsf{N}=3.2$, the TE fundamental is $(m,n)=(2,4)$, Fig.\ \ref{TE32_m1},
however, the $Q$ for this mode is extremely low, so that it cannot be 
considered stable.

\begin{figure}
\includegraphics[angle=-90.0,width=\columnwidth]{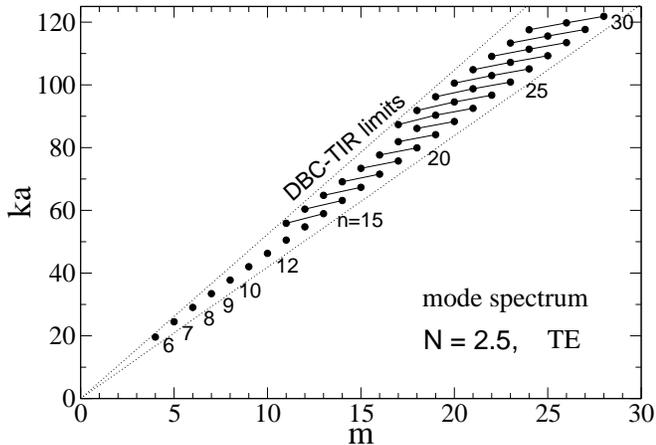}
\caption{ \label{TE2.5_ka} The lowest mode wavevectors for TE polarization
at index ratio $\mathsf{N}=2.5$, plotted as described in Fig.\ \ref{TM3.2_ka}.
}
\end{figure}

In Fig.\ \ref{TE2.5_ka} we illustrate the effect of decreasing
$\mathsf{N}$ down to the value 2.5, closer to the extreme limit 2.0 . 
Two low modes at $m=4$ and $m=5$ have been squeezed out by the lower
DBC-TIR upper limit, as well as the entire spectrum becoming narrower.
Of course, a similar squeezing effect takes places for the TM spectrum.

\begin{figure}
\includegraphics[angle=-90.0,width=\columnwidth]{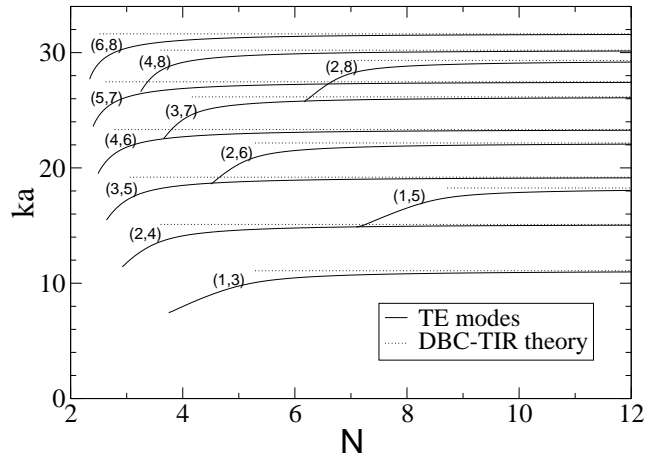}
\caption{ \label{TE-ka-N} TE mode wavevectors (solid curves) as functions of 
the index ratio $\mathsf{N}$, for modes indicated by quantum indexes $(m,n)$.
The dotted lines show the DBC wavevectors, terminating at the cutoffs given
by Eq.\ (\ref{DBC-cut}).
}
\end{figure}

The dependence of some TE mode wavevectors on index ratio is shown in 
Fig.\ \ref{TE-ka-N}.
Compared to the TM modes, the TE modes show a stronger variation
with $\mathsf{N}$, especially just above the cutoff ratio.
The plot also shows the DBC mode wavevectors as dotted lines, terminating
at the cutoffs predicted by the DBC-TIR theory, Eq.\ (\ref{DBC-cut}).
Once $\mathsf{N}$ reaches adequately large values, the wavevectors from 
the calculations for dielectric boundary conditions asymptotically approach 
the DBC values.
This can be attributed to the extra factor of $\epsilon/\epsilon'$ in the 
TE phase shift formula (\ref{deltaTE}), which easily causes all the 
reflection phase shifts to rapidly approach $-\pi$, the value under 
Dirichlet boundary conditions.

\begin{figure}
\includegraphics[angle=0.0,width=\smallfigwidth]{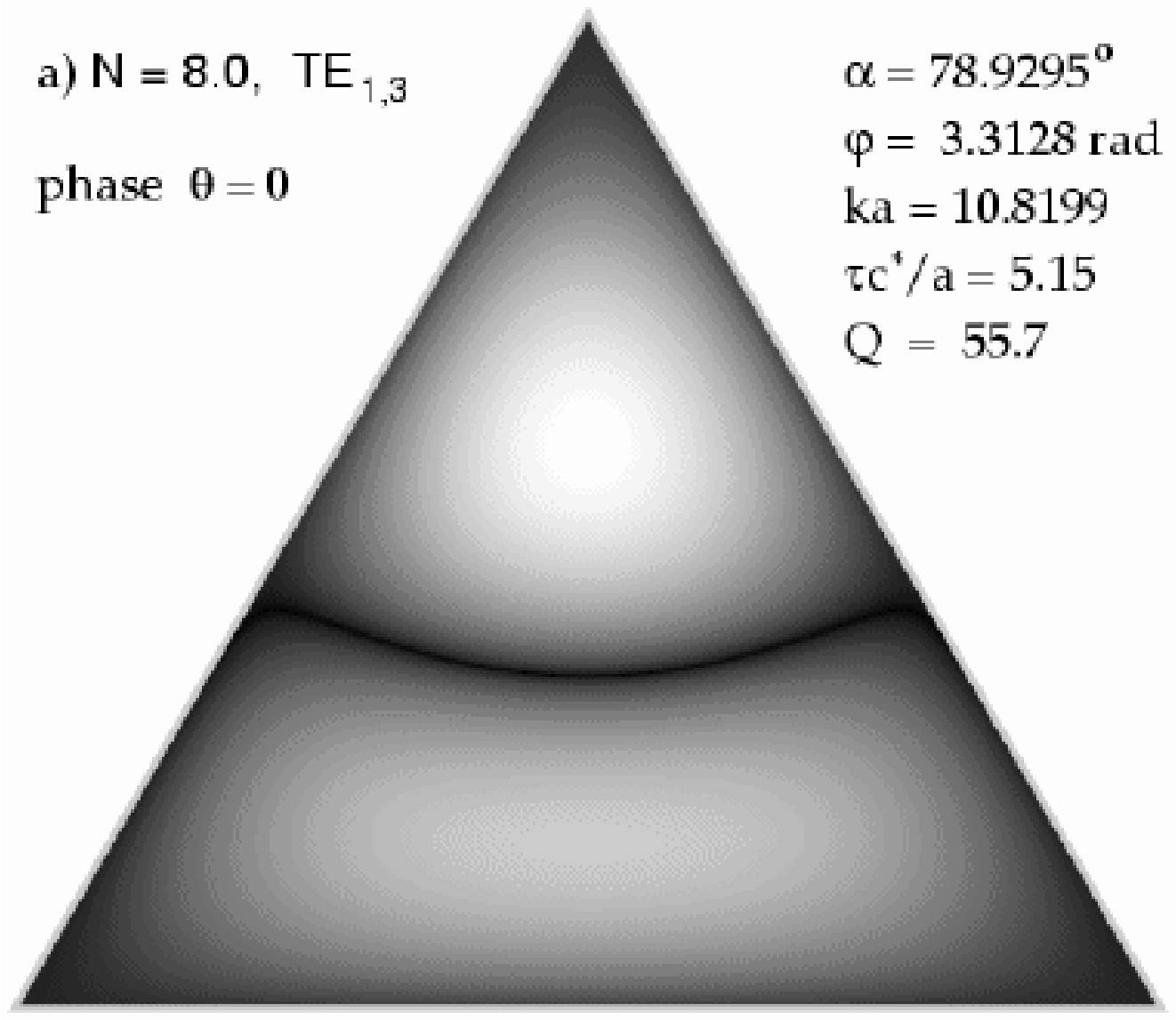}
\includegraphics[angle=0.0,width=\smallfigwidth]{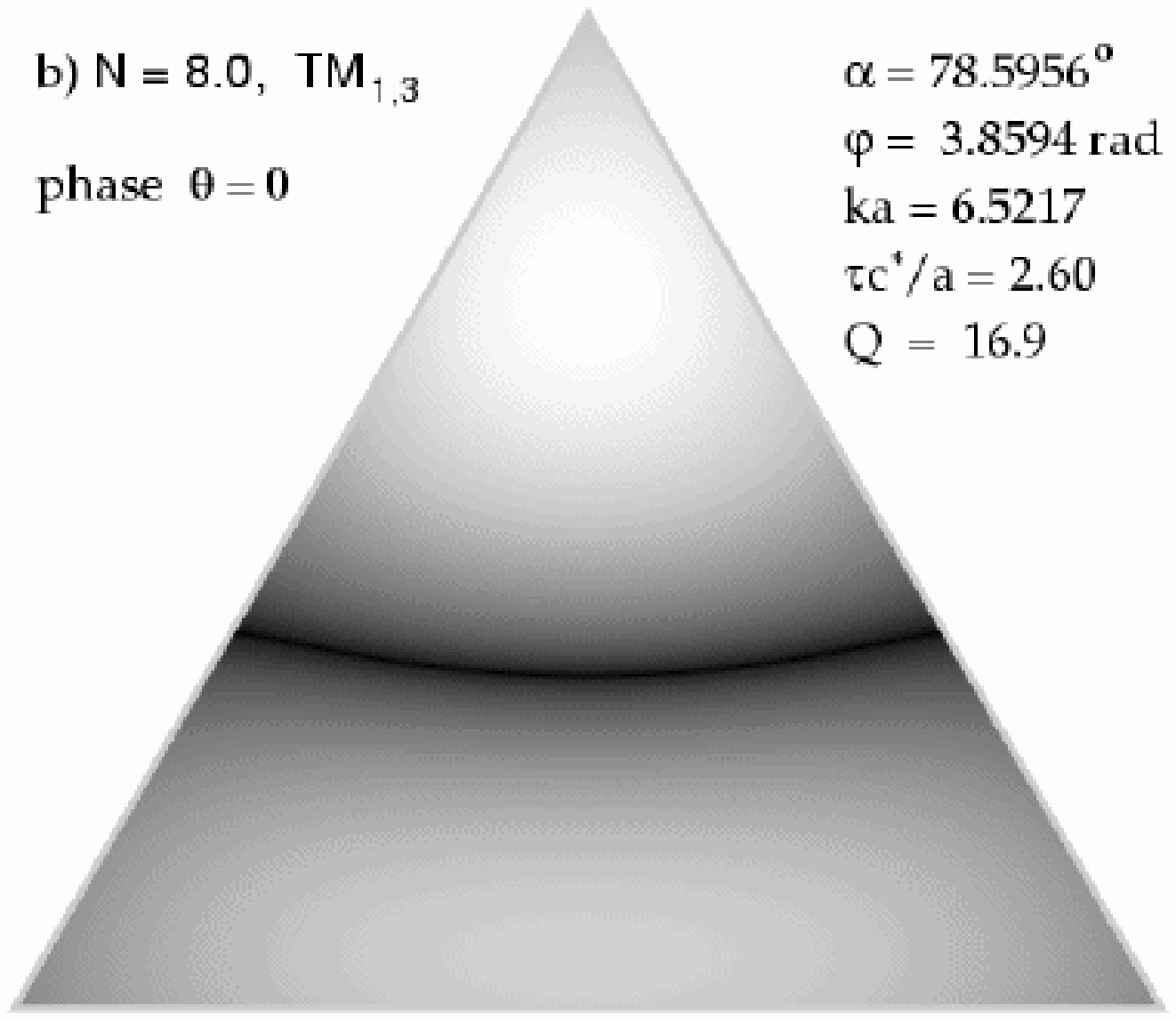}
\caption{ \label{TMTE8_m1}  Fundamental a) TE and b) TM modes 
of oscillation at $\mathsf{N}=8.0$, with $(m,n)=(1,3)$, as described in 
Fig.\ \ref{TM32_m1}, both for phase $\theta=0$, see text.  (The choice
$\theta=\pi/2$ would instead produce a vertical nodal line.)
}
\end{figure}

The fields diminish more rapidly near the cavity edges for TE 
polarization than for TM polarization.
At $\mathsf{N}=8.0$, the fundamental modes have $(m,n)=(1,3)$, whose
corresponding wavefunction is the simplest possible, having a single nodal 
line across the cavity, Fig.\ \ref{TMTE8_m1}.
The avoidance of the fields close to the cavity boundaries of the TE
mode is apparent, whereas, the TM fields tend towards maximum values 
near the boundaries.

Taken together, these results emphatically confirm the idea presented in Ref.\ 
\onlinecite{Wysin05a} that the simple DBC theory is much more appropriate 
for TE polarization than for TM polarization.

\subsection{Estimates of mode lifetimes due to boundary wave emission}
\label{life}
The (doubly-degenerate) solutions found here are approximate.
The assumption of evanescent traveling waves along each boundary,
rotating around the triangle in the exterior region, 
is not exact, because the edges are of finite length.
In reality, any of the evanescent waves can be expected to scatter
from the triangle vertices, leading to radiation away from the cavity,
and reflection of that evanescent wave backwards from the vertices.
There should be linear combinations of evanescent traveling waves 
moving in both directions along the edges (standing waves).
This in turn would lead to a mixing of the doubly degenerate modes,
i.e., the degeneracy will be split due to the scattering experienced
by the fields at the triangular vertices.

In the case of modes whose exterior wavelength is small compared to the
cavity edge, these effects may be small, and ignoring the resulting degeneracy
splitting, we can try to estimate the power loss, and hence, the mode lifetime.
To accomplish this, following Wiersig\cite{Wiersig03}, it is assumed that all
the power in the evanescent boundary waves is radiated when reaching the
vertices of the triangle, without reflecting backwards from the vertices.
%
%
The lifetime $\tau$ and Q-factor are calculated by
\begin{equation}
\label{lifedef}
\tau = U / P, \qquad Q = \omega \tau = 2\pi f \tau,
\end{equation}
where $U$ is the total energy in the cavity fields, and $P$ is the total
power radiated, as found from leakage of the evanescent boundary waves at
the vertices.
This approach was used also to estimate the mode lifetimes using the DBC
solutions\cite{Wysin05a}; there it was found that generally speaking, the TE
modes have longer lifetimes than the corresponding TM modes, provided
the index ratio $\mathsf{N}$ is large.
It is important to check the relative mode lifetimes using the more 
correct dielectric boundary conditions.
As the present solution has demonstrated the avoidance of TE fields
near the cavity boundaries, one might expect the TE lifetimes to be longer.

\begin{figure}
\includegraphics[angle=-90.0,width=\smallfigwidth]{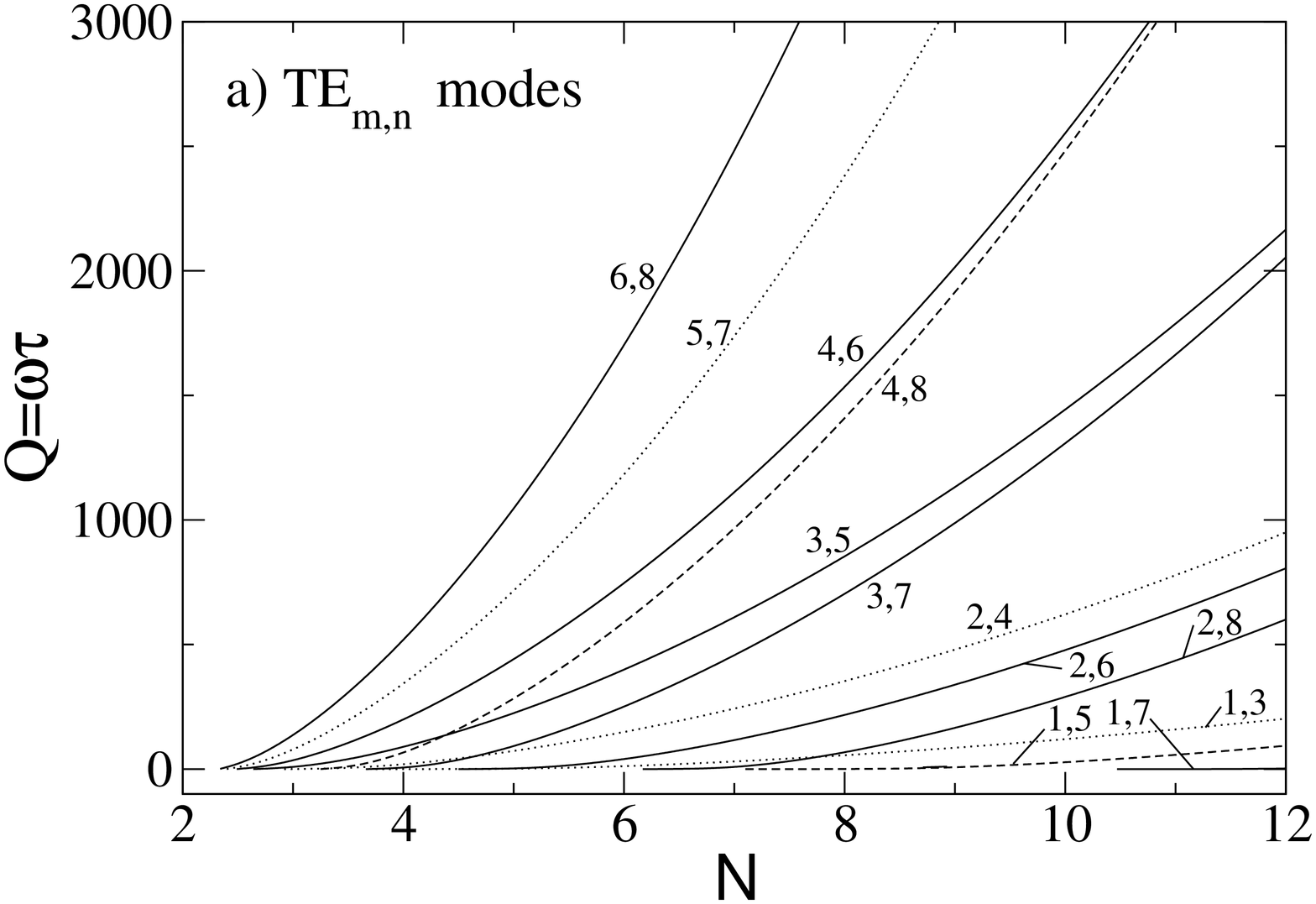}
\includegraphics[angle=-90.0,width=\smallfigwidth]{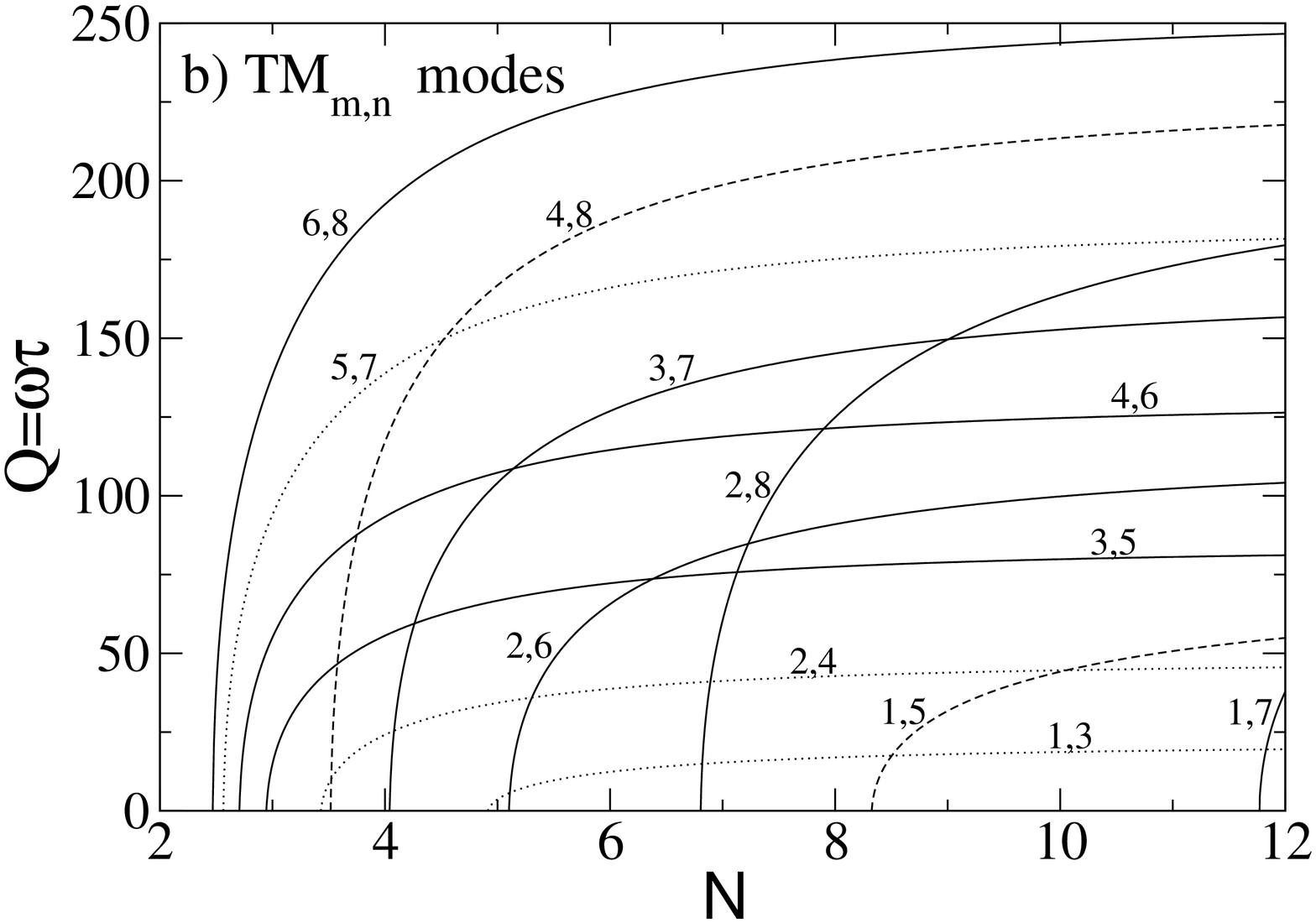}
\caption{ \label{Q_N} Mode quality factors for  a) TE and b) TM modes
calculated from boundary wave emission, see (\ref{tau-formula}) and
(\ref{Q-formula}), as functions of the index ratio $\mathsf{N}$, with modes
labeled by $(m,n)$.
}
\end{figure}

Following the calculations in Refs.\ \onlinecite{Wiersig03,Wysin05a}, 
the total energy of the fields in a cavity of height $h$ can be written
\begin{equation}
\label{U}
U=\int h~ dx ~ dy ~ \frac{\epsilon |\vec{E}|^2}{8\pi}  
 =\int h~ dx ~ dy ~ \frac{\mu      |\vec{H}|^2}{8\pi};
\end{equation}
the two forms convenient for TM modes ($|\vec{E}|=E_z=\psi$) and
TE modes ($|\vec{H}|=H_z=\psi$), respectively.
Using the wavefunction of Eq.\ (\ref{psi}), it is not possible to find
a simple formula for this integral in the general case, thus it was
evaluated by numerical integration within the triangle.
In the DBC limit, with $\varphi=-\pi$, one finds 
$\int \frac{dx}{a} \frac{dy}{a} |\psi|^2 = 3\sqrt{3}|A_0|^2/8$; 
generally also for MBC the integral is of this order.
%

The total boundary wave power is taken as the sum of the powers from boundary
waves on each edge.
At each edge there are three different incident waves, each of which generates
an evanescent wave.
For example,  on $b_0$, waves \textcircled{$\scriptstyle 3$}, 
\textcircled{$\scriptstyle 5$} and \textcircled{$\scriptstyle 6$} separately
produce the evanescent waves \textcircled{$\scriptstyle 3'$},
\textcircled{$\scriptstyle 5'$} and \textcircled{$\scriptstyle 6'$}.
We sum the powers in each of these evanescent waves, then multiply by three 
due to triangular symmetry, to give the total power radiated.

The boundary wave power of an individual evanescent wave is calculated using 
the Poynting flux along the exterior side, 
$\vec{S}' = \frac{c}{8\pi} \mathrm{Re}\{\vec{E}'\times \vec{H}'^{*}\}$,
expressed alternately as
\begin{equation}
\vec{S}' 
= \frac{c}{8\pi\mu'}\vert \vec{E}\;'\vert^2 \mathsf{n}\sin\theta_i \; \hat{x}
= \frac{c}{8\pi\epsilon'}\vert\vec{H}\;'\vert^2 \mathsf{n} \sin\theta_i \;\hat{x},
\end{equation}
where $\hat{x}$ points parallel to the boundary, and the incident angle
from within the cavity is $\theta_i$. 
The evanescent wave has an exponentially decaying behavior into the exterior
medium. 
If $A_i$ is the amplitude on the interior side, with boundary
at $y=0$, and the $y$-coordinate points from the boundary into medium 
$\mathsf{n'}$, Eqs.\ (\ref{A1A6}) and (\ref{gamma}) give
\begin{equation}
|\psi\:'|^2 =  \vert A_i [1+e^{i\delta(\theta_i)}] 
                 e^{ik'_x x} e^{-k'\gamma' y}\vert^2.
\end{equation}
Integrating the total flux contained from $y=0$ to $y=\infty$ gives the power
of this wave along the boundary, 
$P_x = h \int_0^{\infty} dy ~ \vec{S}\:'\cdot\hat{x}$, with only a tiny 
formal difference for the two polarizations,
\begin{equation}
P_x = \frac{hc^2}{4\pi\omega} 
\frac{|A_i|^2 \cos^2[\frac{1}{2}\delta(\theta_i)]}
     {\sqrt{1-(\sin\theta_c/\sin\theta_i)^2}}
\times \left\{ 
\begin{array}{l}
\frac{1}{\mu'} \quad \mathrm{(TM)}, \\
\frac{1}{\epsilon'} \quad \mathrm{(TE)}.
\end{array}
\right.
\end{equation}
In practice, however, the phase shifts for TE polarization are typically
much closer to $-\pi$ than for TM polarization, which causes the TE powers
to be smaller.
The incident squared wave amplitude is $|A_i|^2 = \frac{1}{4}|A_0|^2$, as all 
the six wave components of $\psi$ are of equal strength, Eq.\ (\ref{freely}).

The total emitted power from all edges is three times the sum of the powers 
on edge $b_0$,
\begin{equation}
P = 3(P_{x,3}+P_{x,5}+P_{x,6}).
\end{equation}
Taking the net energy/power ratio and simplifying leads to the 
dimensionless lifetime expressions,
\begin{equation}
\label{tau-formula}
\frac{\tau c^*}{a} = 
\frac{ \frac{2}{3}ka ~ \int \frac{dx}{a}\frac{dy}{a}|\psi|^2 }
  { \sum_{i=3,5,6} \frac{|A_0|^2 \cos^2[\frac{1}{2}\delta_i] }
                        {\sqrt{1-(\sin\theta_c/\sin\theta_i)^2}} }
\times
\left\{
\begin{array}{l}
\frac{\mu'}{\mu} \quad \mathrm{(TM)}, \\
\frac{\epsilon'}{\epsilon} \quad \mathrm{(TE)}.
\end{array}
\right.
\end{equation}
The factor of $\epsilon'/\epsilon < 1$ for TE polarization tends to
reduce the lifetime, however, the phase shifts in the denominator have
an even larger effect, such that typically, the TE$_{m,n}$ lifetime is
found to be longer than the TM$_{m,n}$ lifetime.
The summation of power terms in the denominator is usually dominated
by that of wave \textcircled{$\scriptstyle 6$}, which has the smallest
incident angle.
The expression is scaled by $a/c^*$, the time for the signal to cross
the cavity.
A well-defined mode should have $\tau c^*/a \gg 1$, however, usually it
is more typical to look at the related mode quality factor, $Q$, defined
by
\begin{equation}
\label{Q-formula}
Q = \omega \tau = c^* k \tau = (ka) (\tau c^*/a),
\end{equation}
whihc is the dimensionless mode wavevector times the
dimensionless lifetime.

\begin{figure}
\includegraphics[angle=0.0,width=\smallfigwidth]{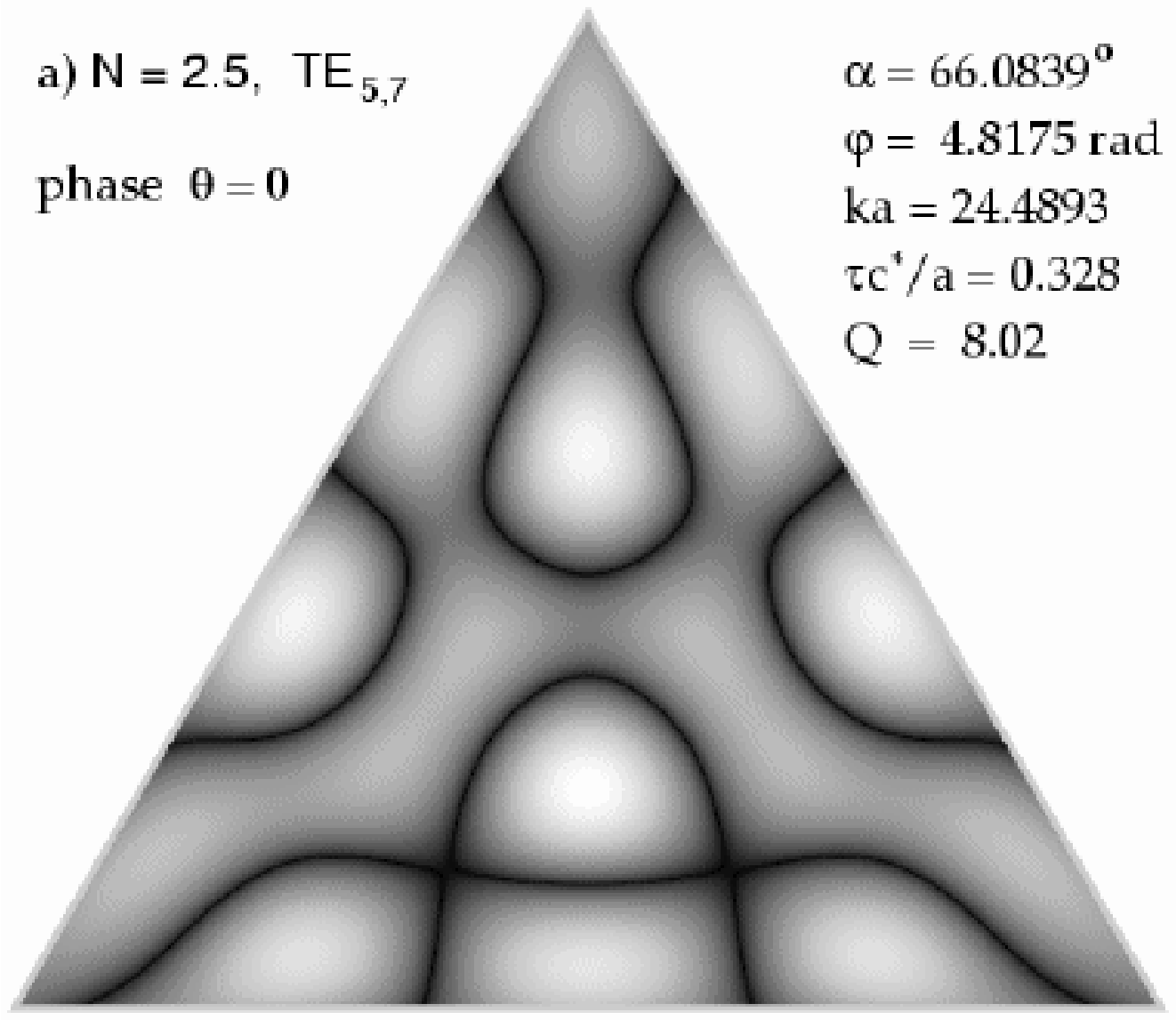}
\includegraphics[angle=0.0,width=\smallfigwidth]{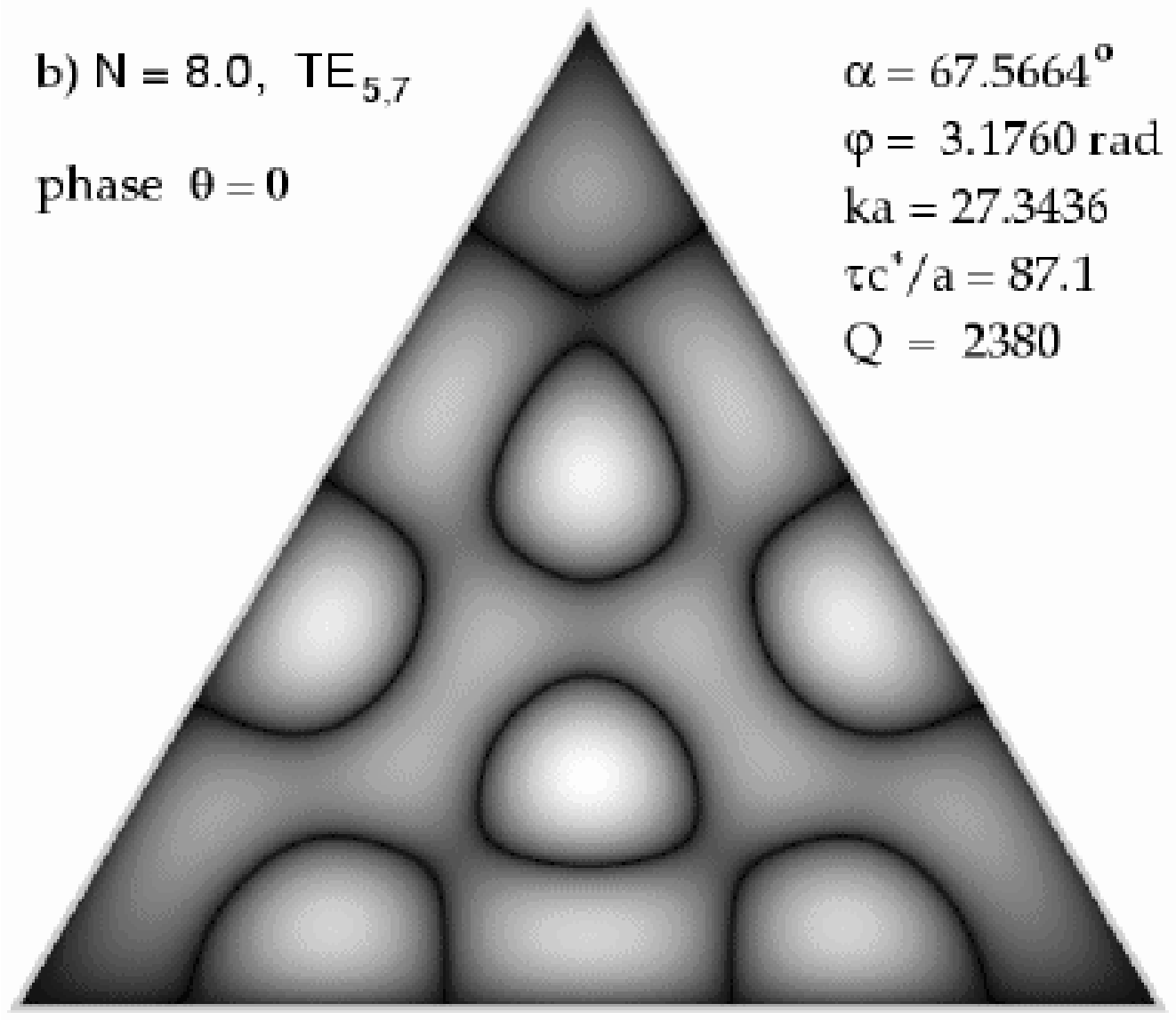}
\caption{ \label{m5_7}  Comparison of TE$_{5,7}$ wavefunctions at
 a) $\mathsf{N}=2.5$, only slightly above cutoff, and very weakly bound with 
small $Q$ and b) $\mathsf{N}=8.0$, substantially above the cutoff, with
much higher $Q$, and field amplitude concentrated away from the boundaries.
}
\end{figure}

Numerical results for $Q$ versus index ratio are displayed for the lowest
modes in Fig.\ \ref{Q_N}, for TE and TM polarizations, assuming $\mu=\mu'$.
One finds that $Q$'s for TE modes are considerably larger than for TM
modes, especially far enough above the cutoff $\mathsf{N}$ for a
given mode.
Furthermore, the TM quality factors tend to saturate at large $\mathsf{N}$,
while the TE quality factors tend to increase proportional to 
$\mathsf{N}^2 = \epsilon/\epsilon'$.
This latter effect can be seen due to the asymptotics for the TE reflection
phase shift, based on the identity,
\begin{equation}
\cos^2\frac{\delta}{2} = \left(\frac{\epsilon'}{\epsilon}\right)^2
\frac{\cos^2 \theta_i}{\sin^2 \theta_i -\sin^2\theta_c +(\epsilon'/\epsilon)^2
\cos^2 \theta_i }.
\end{equation}
The net result is that $Q$ and $\tau$ for TE modes are proportional to
$\epsilon/\epsilon'$ in the limit of large index ratio, a factor not 
present for TM polarization.

\begin{figure}
\includegraphics[angle=-90.0,width=\smallfigwidth]{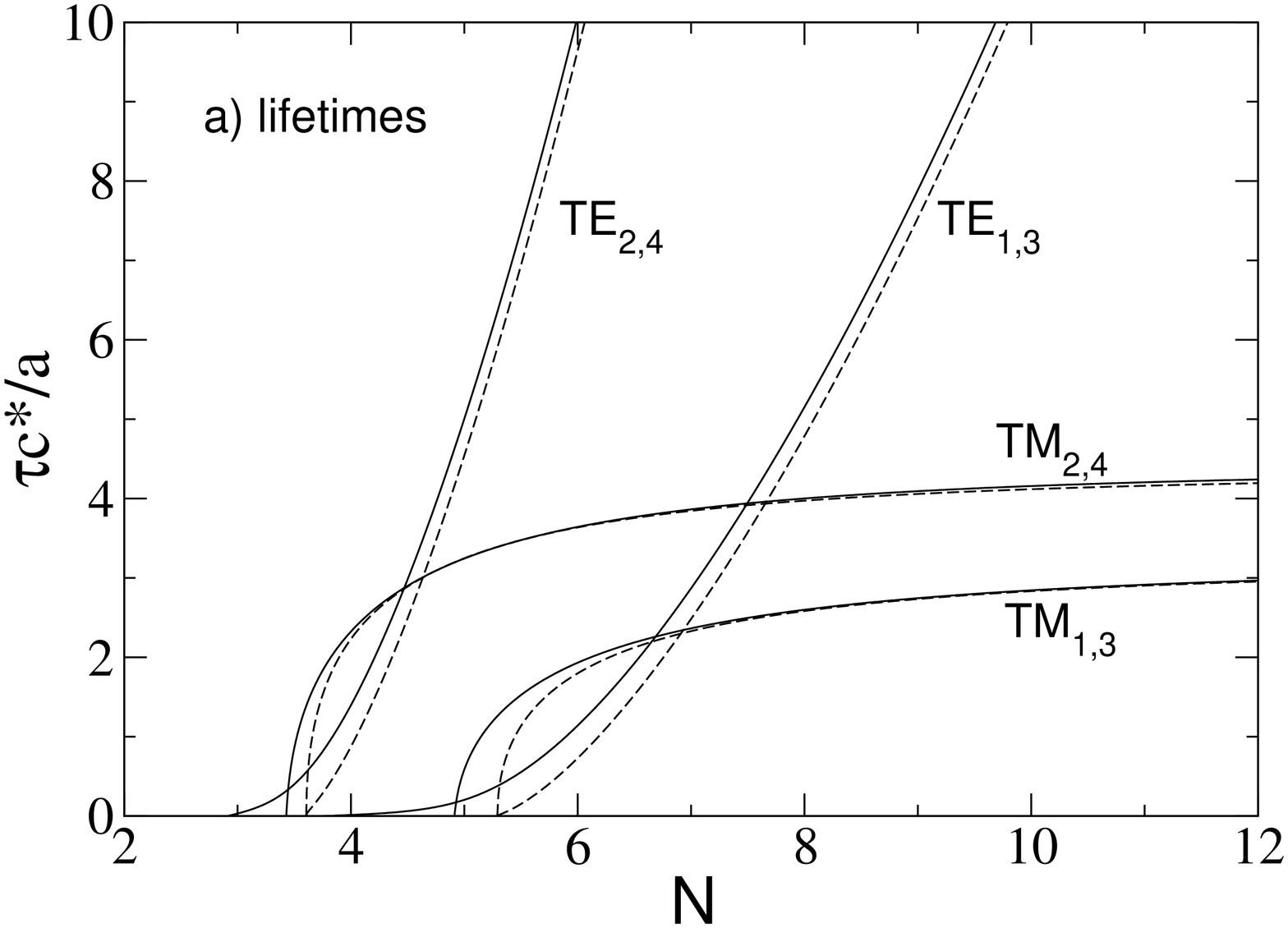}
\includegraphics[angle=-90.0,width=\smallfigwidth]{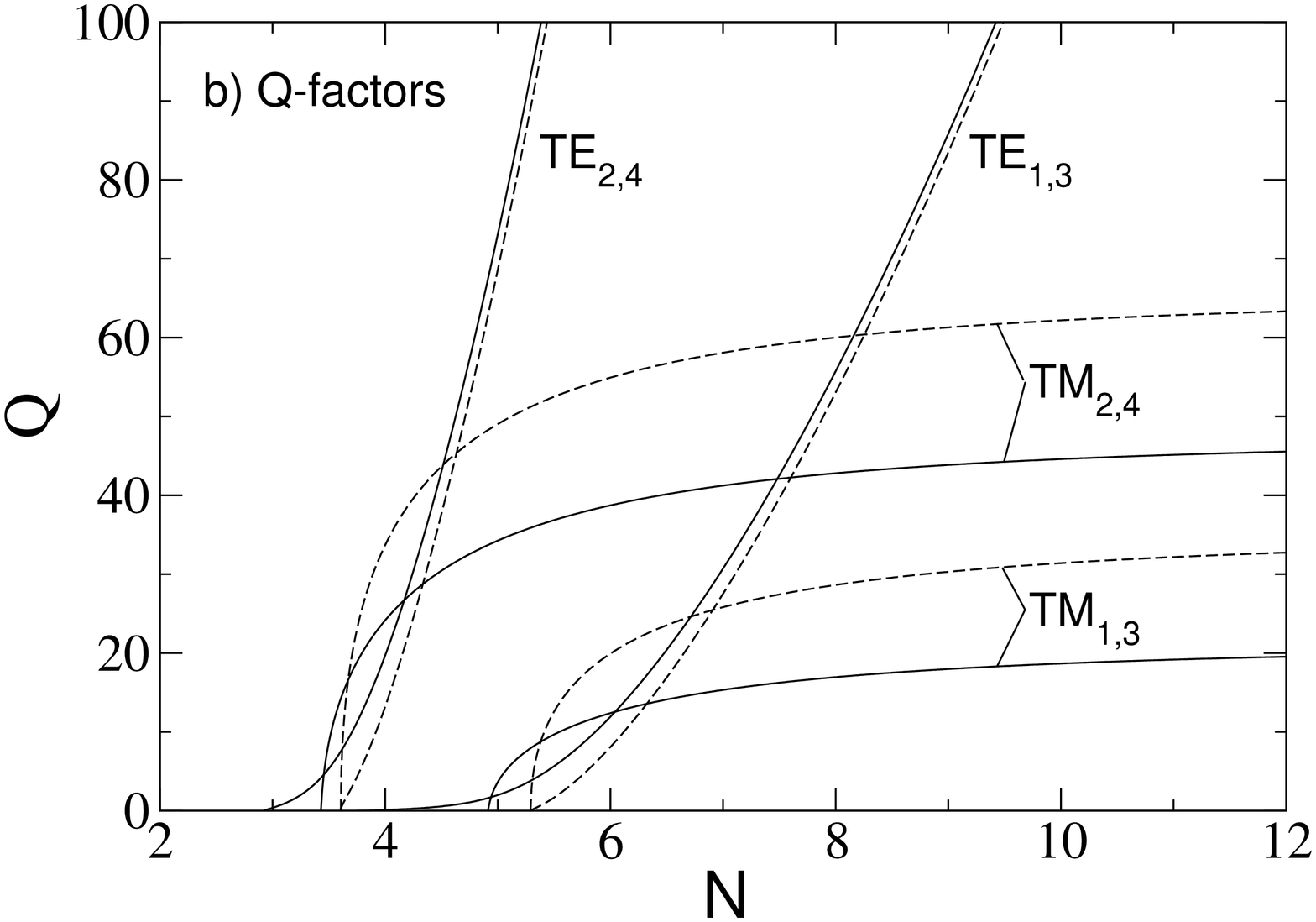}
\caption{ \label{tau_compare} Comparison of some mode a) lifetimes
and b) quality factors, as functions of the index ratio $\mathsf{N}$, 
from the present theory, using dielectric boundary conditions 
(solid curves), and from the simpler theory using Dirichlet 
boundary conditions (dashed curves). The modes are labeled by $(m,n)$.
}
\end{figure}

It is interesting to observe the changes in the wavefunctions with
increasing $\mathsf{N}$ or effectively, increasing $Q$.
In Fig.\ \ref{m5_7},  TE$_{5,7}$ mode wavefunctions are plotted for
$\mathsf{N}=2.5$, just slightly above the cutoff index ratio, and
for $\mathsf{N}=8$, well above the cutoff.
At lower $\mathsf{N}$ (and $Q$), a central lobe of $\psi$ is connected
to one vertex of the boundary, and another lobe is nearly coupled
to the opposite edge.
At higher $\mathsf{N}$ (and much higher $Q$), these interior lobes have
become completely detached from the boundary, and the fields appear more
concentrated within the interior, farther from the boundaries.

A comparison can be made between the results using dielectric boundary 
conditions and those from the simplified DBC theory\cite{Wysin05a}.
Fig.\ \ref{tau_compare} shows $Q$ and $\tau$ results for the modes 
$(1,3)$ and $(2,4)$, as derived from the two approaches, using the 
total power in all boundary waves for this comparison.
Indeed, there are only minor differences between the lifetimes, 
mostly near the cutoff index ratios, due to the fact that
the DBC theory overestimates the cutoffs.
For the quality factors, the discrepancy between the approaches is
much greater for the TM modes, primarily because the DBC theory has
overestimated their frequencies (or wavevectors).
Nevertheless, the ability of the simplified DBC theory to predict the
general trends with $\mathsf{N}$ is remarkable.

\begin{table}
\caption{\label{loTMmodes} Properties of some of the lower TM mode 
for $\mathsf{N}=3.2$, labeled by $(m,n)$ or by 
underlined indexes $(\underline{m},\underline{l})$, where
compared to results of Ref.\ \onlinecite{Huang01}, in parenthesis.
}
\begin{ruledtabular}
\begin{tabular}{cccc}
  mode & $\alpha$ &  $ka$  &  $Q$ \\
\hline
TM$_{3,5}$ & 70.4167$^\circ$ & 14.3406  & 32.18 \\
TM$_{4,6~(\underline{0},\underline{14})}$ 
   & 68.6412$^\circ$ & 18.5733~(18.8)  & 67.67 \\
TM$_{5,7~(\underline{0},\underline{16})}$ 
   & 67.3705$^\circ$ & 22.7686~(22.9)   & 107.9 \\
TM$_{6,8~(\underline{0},\underline{18})}$ 
   & 66.4225$^\circ$ & 26.9545~(27.0)   & 154.5 \\
TM$_{7,9~(\underline{0},\underline{20})}$ 
    & 65.6894$^\circ$ & 31.1374~(31.2)  & 207.9 \\
TM$_{8,10~(\underline{0},\underline{22})}$ 
    & 65.1060$^\circ$ & 35.3196~(35.3)  & 268.4 \\
TM$_{6,10~(\underline{1},\underline{21})}$ 
    & 70.6463$^\circ$ & 33.5168~(34.1)  & 145.8 \\
TM$_{9,11~(\underline{0},\underline{24})}$ 
    & 64.6307$^\circ$ & 39.5019~(39.5)  & 335.9 \\
TM$_{7,11~(\underline{1},\underline{23})}$ 
    & 69.6339$^\circ$ & 37.7112~(38.3)  & 227.4 \\
TM$_{10,12}$ & 64.2363$^\circ$ & 43.6845 & 410.4 \\
TM$_{8,12}$ & 68.7916$^\circ$ & 41.8887 & 307.3 \\
 $\ldots$ & $\ldots$ &  $\ldots$ & $\ldots$  \\
TM$_{15,17~(\underline{0},\underline{36})}$ 
    & 62.9701$^\circ$ & 64.6045~(64.5) & 889.9~(15000) \\
TM$_{13,17~(\underline{1},\underline{35})}$ 
    & 66.1053$^\circ$ & 62.7505~(62.9) & 784.1~(4380) \\
TM$_{11,17}$ & 69.3950$^\circ$ & 61.0455 & 593.5 \\
TM$_{16,18~(\underline{0},\underline{38})}$ 
    & 62.8025$^\circ$ & 68.7897~(68.7) & 1007~(22000) \\
TM$_{14,18~(\underline{1},\underline{37})}$ 
    & 65.7527$^\circ$ & 66.9250~(67.0) & 899.0~(7000) \\
TM$_{12,18~(\underline{2},\underline{36})}$ 
    & 68.8431$^\circ$ & 65.2085~(65.5) & 717.7~(1230) \\
TM$_{17,19}$ & 62.6528$^\circ$ & 72.9751 & 1132 \\
TM$_{15,19}$ & 65.4386$^\circ$ & 71.1006 & 1021 \\
TM$_{13,19~(\underline{2},\underline{38})}$ 
    & 68.3516$^\circ$ & 69.3713~(69.5) & 845.1~(1960) \\
\end{tabular}
\end{ruledtabular}
\end{table}

\begin{table}
\caption{\label{loTEmodes} Properties of some of the lower TE modes 
for $\mathsf{N}=3.2$, labeled by $(m,n)$ or by 
underlined indexes $(\underline{m},\underline{l})$, where
compared to results of Ref.\ \onlinecite{Huang01}, in parenthesis.
}
\begin{ruledtabular}
\begin{tabular}{cccc}
  mode & $\alpha$ &  $ka$  &  $Q$ \\
\hline
TE$_{2,4}$ & 71.5207$^\circ$ & 12.5628 & 1.576 \\
TE$_{3,5}$ & 69.9852$^\circ$ & 17.6896 & 18.95 \\
TE$_{4,6}$ & 68.4450$^\circ$ & 22.1165 & 60.13 \\
TE$_{5,7~(\underline{0},\underline{16})}$ 
    & 67.2586$^\circ$ & 26.3873~(26.2) & 121.4 \\
TE$_{6,8}$ & 66.3498$^\circ$ & 30.6105 & 200.1 \\
TE$_{7,9}$ & 65.6382$^\circ$ & 34.8149 & 294.6 \\
TE$_{5,9}$ & 71.3190$^\circ$ & 32.0399 & 15.63 \\
TE$_{8,10~(\underline{0},\underline{22})}$ 
     & 65.0678$^\circ$ & 39.0107~(39.0) & 404.1 \\
TE$_{6,10}$ & 70.3771$^\circ$ & 36.7552 & 59.80 \\
 $\ldots$ & $\ldots$ &  $\ldots$ & $\ldots$  \\
TE$_{14,16~(\underline{0},\underline{34})}$
     & 63.1471$^\circ$ & 64.1404~(64.2) & 1350~(6130) \\
TE$_{12,16~(\underline{1},\underline{33})}$ 
     & 66.4654$^\circ$ & 62.2279~(62.4) & 789~(4100) \\
TE$_{10,16}$ & 69.8843$^\circ$ & 60.2288 & 221 \\
TE$_{15,17~(\underline{0},\underline{36}}$ 
     & 62.9597$^\circ$ & 68.3271~(68.4) & 1560~(10860) \\
TE$_{13,17~(\underline{1},\underline{35})}$
     & 66.0736$^\circ$ & 66.4147~(66.6) & 972~(15320) \\
TE$_{11,17}$ & 69.2967$^\circ$ & 64.4910 & 346 \\
\end{tabular}
\end{ruledtabular}
\end{table}

\subsection{Comparison with other ETR theory for dielectric 
boundary conditions}
ETR modes have previously been analyzed by Huang 
\textit{et al.}(HGW)\cite{Huang01} using different approximations, also 
involving matching of interior fields undergoing TIR to exterior 
evanescent fields.
In particular, HGW used what was called a ``perfectly confined
approximation for the transverse wavefunction.''
This is a Dirichlet boundary condition on part of the full wavefunction, 
giving the $y$-component of wavevector \textcircled{$\scriptstyle 1$}
[Eq.\ (9) of Ref.\ \onlinecite{Huang01}],
\begin{equation} 
k_y a = \frac{2\pi}{\sqrt{3}}(\underline{m}+1), 
\qquad \underline{m}=0,1,2, \ldots
\end{equation}
Indeed, this is the same as Eq.\ (\ref{kxkyDBC}) reviewed here for the DBC 
problem, with $\underline{m}+1$ equivalent to $n=n_6+2$.
Thus, it is noticeably different from the result for dielectric boundary 
conditions, Eq.\ (\ref{kxky}b), which more fully accounts for all the 
reflection phase shifts.
The factor $\underline{m}+1$ begins at the value 1, whereas, in our 
results, the minimum value of equivalent quantum index $n$ is 3.
For the other component of wavevector \textcircled{$\scriptstyle 1$},
the result obtained [Eq.\ (21) of Ref.\ \onlinecite{Huang01}] is similar to
Eq.\ (\ref{kxky}a),
\begin{equation}
k_x a = \frac{2\pi}{3} \underline{l} -2\underline{\theta}, 
\qquad \underline{l}=3,4,5,\ldots
\end{equation}
where $\underline{l}$ has the same parity as $\underline{m}$, and 
$\underline{\theta}$ is an individual TIR Fresnel phase shift, having 
different forms for TM and TE modes.
Index $\underline{l}$ appears equivalent to $m=n_3-n_5$, however, Eq.\
(\ref{kxky}a) involves phase shifts of two different waves, instead of
the individual phase shift $\underline{\theta}$.

For a chosen value of $n$, different choices of $m$ give solutions 
with $ka$ of similar magnitude, like the modes connected by solid lines 
in Figs.\ \ref{TM3.2_ka}, \ref{TM8.0_ka}, \ref{TE3.2_ka} and \ref{TE2.5_ka}.
This ``number of transverse modes'' increases with both $n$ and $\mathsf{N}$,
and is smaller than that found in Ref.\ \onlinecite{Huang01}.
Because $ka=2\pi\mathsf{N}/\lambda$ depends on $(m,n)$ in a nontrivial way,
it is not possible to give a simple expression for this mode count.

For $\mathsf{N}=3.2$, a summary is made of the lowest TM and TE modes
in Tables \ref{loTMmodes} and \ref{loTEmodes} and compared to results from HGW.
In spite of the obvious differences in these theoretical approaches, they 
both lead to very similar predictions for the mode wavevectors or frequencies,
agreeing to within about one percent.
The prediction of the $Q$-values are considerably different;
HGW used the finite-difference time-domain (FDTD)
technique\cite{Dey98} combined with Pad\'e approximates to estimate $Q$.
In particular, the FDTD technique predicts that TE polarization results
in much smaller $Q$ than TM polarization\cite{Guo00}, exactly opposite to 
our results (Fig.\ \ref{Q_N}).
The short lifetime for TE modes was explained by the zero in the reflectivity
at the Brewster angle $\theta_B$, which is only slightly less than the 
critical angle.
However, in a confined TIR wavefunction, all six plane wave components 
must impinge on the faces at angles greater than $\theta_c$ and hence 
greater than $\theta_B$, so it is hard to understand how the Brewster
minimum comes into play, unless diffractive effects at the triangle
vertices are strong, causing violation of the six-wave assumption.

\subsection{Mode generation near 1.55 $\mu$m free space wavelength}
There has been recent interest, both experimental and theoretical,
in using semiconductor ETRs operating around 1.55 $\mu$ m wavelength.
\cite{Guo00,Huang01,Huang+01,Lu04}
Here we assume a cavity medium with $\mathsf{n}=3.2$, surrounded by
vacuum, and summarize the mode spectra obtained by the present theory,
for some typical cavity edge lengths, $a=2 \mu$m and  $a=5 \mu$m.
Of course, all mode frequencies simply scale as $1/a$, thereby giving
the primary tuning control parameter.
Modes whose wavelength outside the cavity ranging from approximately
$1.30 \mu$m to $1.60 \mu$m are considered; 
this corresponds to frequency range 187 THz $ < f < $ 231 THz.

\begin{table}
\caption{\label{TMTE32modes} Mode frequencies and free space wavelengths 
in the range around 1.3 to 1.6 $\mu$m, for cavities with $\mathsf{N}=3.2$ 
and edge lengths $a$. The modes are labeled by $m,n$.  Wavelengths marked
with an asterisk fall about 2\% below peaks in PL data, Ref.\ \onlinecite{Lu04}
}
\begin{ruledtabular}
\begin{tabular}{ccccc}
 $a$ ($\mu$m)  & mode  & $ka$ & $\lambda$ ($\mu$m) & $f$ (THz) \\
\hline
2 & TM$_{3,5}$ & 14.341 & 1.5675 & 191.25 \\
2 & TE$_{1,3}$ & 12.563 & 1.7893 & 167.54 \\
\hline
5 & TM$_{6,10}$ & 33.517 & 1.6767 & 178.79 \\
5 & TM$_{8,10}$ & 35.320 & 1.5911* & 188.41 \\
5 & TM$_{7,11}$ & 37.711 & 1.4902* & 201.17 \\
5 & TM$_{9,11}$ & 39.502 & 1.4227* & 210.72 \\
5 & TM$_{8,12}$ & 41.889 & 1.3416* & 223.46 \\
5 & TM$_{10,12}$ & 43.685 & 1.2865* & 233.03 \\
5 & TE$_{5,9}$ & 32.040 & 1.7540 & 170.92 \\
5 & TE$_{7,9}$ & 34.815 & 1.6142 & 185.72 \\
5 & TE$_{6,10}$ & 36.755 & 1.5290 & 196.07 \\
5 & TE$_{8,10}$ & 39.011 & 1.4406 & 208.10 \\
5 & TE$_{7,11}$ & 41.135 & 1.3662 & 219.43 \\
5 & TE$_{9,11}$ & 43.202 & 1.3008 & 230.46 \\
\end{tabular}
\end{ruledtabular}
\end{table}

Results for all the modes found from 1.3 to 1.6 $\mu$m are shown in Table
\ref{TMTE32modes} for TM and TE polarization.
Obviously very few modes occur in any such narrow range if the cavity
is small ($a=2\mu$m), which could allow for fine frequency tuning.
Conversely, many modes are present in larger cavities
(such as $a= 10 \mu$m) and single mode operation is difficult.
A cavity with $a=5 \mu$m has a moderate number of modes; the TM mode 
wavelengths are similar to those found in photoluminescence (PL)
experiments\cite{Lu04}.
In fact, five of TM mode wavelengths calculated here are about 2\% 
lower than five peaks seen in Fig.\ 2 of Ref. \onlinecite{Lu04}.
This deviation might be attributed primarily to an uncertainty in the cavity
size, and secondly, caused by a weak variation of dielectric constant with 
wavelength.\cite{Lu04}
Discounting these factors, the agreement for the $5 \mu$m cavity is reasonable.
Comparison with the experimental spectrum for $a= 10\mu$m is more difficult,
although it appears that some of the TM$_{m,m+2}$ modes from the present theory
(not shown) do appear in the PL data, again allowing for uncertainty in $a$ 
and dispersion.

\section{Conclusions}
The phase relationships between the six plane waves within an ETR have been 
determined so they match correctly to each other and to exterior evanescent
waves,  according to Fresnel reflection coefficients for a dielectric on 
dielectric boundary at index ratio $\mathsf{N=n/n'}$.
The theoretical wavefunction description is actually very general; it 
applies to any choice of reflection phase shifts, including those for 
DBC, and TE or TM polarizations.
The main approximation here has been that the evanescent fields do not
perturb the interior fields;  assuming the evanescent fields radiate at
the triangle vertices (a strong damping approximation), the mode lifetimes
and quality factors have been estimated.
Modes with high $Q$ should be very well described by the wavefunction 
(\ref{psi}).
The mode wavevectors are consistent with previous analyses
\cite{Guo00,Huang01,Huang+01}, although the wavefunction description 
and quality factors differ from the FDTD technique\cite{Guo00} results.
Some mode results are consistent with PL experimental data on 
$5 \mu$m semiconductor ETRs\cite{Lu04}.

For $\mathsf{N}$ adequately above the cutoff for a mode $(m,n)$, the TE mode
wavevectors are very close to the predictions from the simplified DBC
theory, whereas, the TM mode wavevectors are consistently below the DBC
results.
This is because large $\mathsf{N}$ results in Fresnel reflection phase shifts
very near $-\pi$, the value for the DBC theory, only for TE polarization.
This causes the TE wavefunctions to avoid the cavity edges; on the other
hand, the TM wavefunctions have significant amplitude at the edges.
For both polarizations, dielectric boundary conditions give lower
cutoff index ratios than from the DBC theory.
As $\mathsf{N}$ increases starting from 2.0, the theory demonstrates how
the spectrum of available modes expands (Fig.\ \ref{TM3.2_ka}, etc.),
but always stays within the limits predicted from the DBC theory.

An extra factor of $\epsilon/\epsilon'$ appears in the lifetime for TE
modes, not present for TM modes.
This theory then predicts considerably larger lifetimes and $Q$'s for
TE polarization, whereas the FDTD approach\cite{Guo00} led to larger $Q$s
for TM modes.
At large index ratio, the lifetimes found here approach the values found 
in the simpler DBC theory\cite{Wysin05a}, for both polarizations.

\begin{acknowledgments}
The author is grateful for discussions with Wagner Figueiredo and 
Luis G. C. Rego and their hospitality at the Universidade Federal de Santa 
Catarina, Florian\'opolis, Brazil, where this work was completed.
\end{acknowledgments}

\bibliography{waveoptics,wysin}

\end{document}